\documentclass[longauth]{aa}  

\usepackage{graphicx}
\usepackage{amsmath}
\usepackage{txfonts}
\usepackage{xcolor}
\usepackage{lipsum}
\usepackage{subcaption}         
\usepackage{lscape}             
\usepackage{placeins}           

\usepackage{hyperref}    
\graphicspath{ {Figures/} }
                                
\begin{document}

   \title{The GAPS programme at TNG
   }

   \subtitle{LXXVI. TOI-1533: A compact system hosting a super-Neptune-mass pair with disparate radii}

   \author{G. Mantovan
          \inst{\ref{inst1},\ref{inst2}}$^{\orcid{0000-0002-6871-6131}}$
          \and 
          V. Nascimbeni\inst{\ref{inst2}}$^{\orcid{0000-0001-9770-1214}}$
          \and
          S. Desidera\inst{\ref{inst2}}$^{\orcid{0000-0001-8613-2589}}$
          \and
          L. Malavolta\inst{\ref{inst3},\ref{inst2}}$^{\orcid{0000-0002-6492-2085}}$
          \and
          J. J. Lissauer\inst{\ref{inst4}}$^{\orcid{0000-0001-6513-1659}}$
          \and
          P. Leonardi\inst{\ref{inst3},\ref{inst2}}$^{\orcid{0000-0001-6026-9202}}$
          \and
          T. Azevedo Silva\inst{\ref{inst5}}$^{\orcid{0000-0002-9379-4895}}$
          \and
          C. Guerra\inst{\ref{inst6}}
          \and
          D. Polychroni\inst{\ref{inst7}}$^{\orcid{0000-0002-7657-7418}}$
          \and
          L. Borsato\inst{\ref{inst2}}$^{\orcid{0000-0003-0066-9268}}$
          \and
          M. Baratella\inst{\ref{inst8}}$^{\orcid{0000-0002-1027-5003}}$
          \and
          K. Biazzo\inst{\ref{inst9}}$^{\orcid{0000-0002-1892-2180}}$
          \and
          D. Nardiello\inst{\ref{inst3},\ref{inst2}}$^{\orcid{0000-0003-1149-3659}}$
          \and
          K. A. Collins\inst{\ref{inst10}}$^{\orcid{0000-0001-6588-9574}}$
          \and
          M. Damasso\inst{\ref{inst11}}$^{\orcid{0000-0001-9984-4278}}$
          \and
          J. De Leon\inst{\ref{inst12}}$^{\orcid{0000-0002-6424-3410}}$
          \and
          M. E. Everett \inst{\ref{inst13}}$^{\orcid{0000-0002-0885-7215}}$
          \and
          D. Gandolfi\inst{\ref{inst14}}$^{\orcid{0000-0001-8627-9628}}$
          \and
          S. Giacalone\inst{\ref{inst15}}$^{\orcid{0000-0002-8965-3969}}$
          \and
          L. Naponiello\inst{\ref{inst11}}$^{\orcid{0000-0001-9390-0988}}$
          \and
          G. Piotto\inst{\ref{inst3},\ref{inst2}}$^{\orcid{0000-0002-9937-6387}}$
          \and
          G. Scandariato\inst{\ref{inst16}}$^{\orcid{0000-0003-2029-0626}}$
          \and
          K. Stassun\inst{\ref{inst17}}$^{\orcid{0000-0002-3481-9052}}$
          \and
          S.~W.~Yee \inst{\ref{inst18}, \footnote{51 Pegasi b Fellow}}$^{\orcid{0000-0001-7961-3907}}$
          \and
          L. Affer\inst{\ref{inst6}}$^{\orcid{0000-0001-5600-3778}}$
          \and
          F. Amadori\inst{\ref{inst11}}$^{\orcid{0000-0003-1316-1033}}$
          \and
          A. Bignamini\inst{\ref{inst7}}$^{\orcid{0000-0002-5606-6354}}$
          \and
          S. Colombo\inst{\ref{inst6}}$^{\orcid{0000-0002-3257-862X}}$
          \and
          M. D'Arpa\inst{\ref{inst6}}$^{\orcid{0009-0004-5914-7274}}$
          \and
          C. D. Dressing\inst{\ref{inst19}}$^{\orcid{0000-0001-8189-0233}}$
          \and
          A. Fukui\inst{\ref{inst12},\ref{inst20}}$^{\orcid{0000-0002-4909-5763}}$
          \and
          A. Ghedina\inst{\ref{inst21}}$^{\orcid{0000-0003-4702-5152}}$
          \and
          J. Higuera\inst{\ref{inst13}}$^{\orcid{0000-0002-3985-8528}}$
          \and
          K. Ikuta\inst{\ref{inst22}}$^{\orcid{0000-0002-5978-057X}}$
          \and
          Kawauchi\inst{\ref{inst23}}$^{\orcid{0000-0003-1205-5108}}$
          \and
          J. Korth\inst{\ref{inst24}, \ref{inst20}, \ref{inst25}}$^{\orcid{0000-0002-0076-6239}}$
          \and
          V. Lorenzi\inst{\ref{inst21}}$^{\orcid{0000-0002-1958-9930}}$
          \and
          L. Mancini\inst{\ref{inst26},\ref{inst11},\ref{inst27}}$^{\orcid{0000-0002-9428-8732}}$
          \and
          M. Mori\inst{\ref{inst28},\ref{inst29}}$^{\orcid{0000-0003-1368-6593}}$
          \and
          F. Murgas\inst{\ref{inst20},\ref{inst25}}$^{\orcid{0000-0001-9087-1245}}$
          \and
          N. Narita\inst{\ref{inst12},\ref{inst28},\ref{inst20}}$^{\orcid{0000-0001-8511-2981}}$
          \and
          E. Palle\inst{\ref{inst20},\ref{inst25}}$^{\orcid{0000-0003-0987-1593}}$
          \and
          H. Parviainen\inst{\ref{inst20},\ref{inst25}}$^{\orcid{0000-0001-5519-1391}}$
          \and
          A. Ruggieri\inst{\ref{inst2}}$^{\orcid{0000-0001-9349-1272}}$
          \and
          A. Savel\inst{\ref{inst30}}$^{\orcid{0000-0002-2454-768X}}$
          \and
          R. P. Schwarz\inst{\ref{inst10}}$^{\orcid{0000-0001-8227-1020}}$
          \and
          A. Shporer\inst{\ref{inst31}}$^{\orcid{0000-0002-1836-3120}}$
          \and
          G. Srdoc\inst{\ref{inst32}}
          \and
          Stockdale\inst{\ref{inst33}}$^{\orcid{0000-0003-2163-1437}}$
          \and
          D. Watanabe\inst{\ref{inst34}}
          \and
          R. Zambelli\inst{\ref{inst35}}
          \and
          T. Zingales\inst{\ref{inst3},\ref{inst2}}$^{\orcid{0000-0001-6880-5356}}$
        }

   \institute{Centro di Ateneo di Studi e Attivit\`a Spaziali ``G. Colombo'' -- Universit\`a di Padova, Via Venezia 15, IT-35131, Padova, Italy; \email{giacomo.mantovan@unipd.it}\label{inst1}
            \and
             Istituto Nazionale di Astrofisica - Osservatorio Astronomico di Padova, Vicolo dell'Osservatorio 5, IT-35122, Padova, Italy\label{inst2}
             \and
             Dipartimento di Fisica e Astronomia ``Galileo Galilei'', Università di Padova, Vicolo dell'Osservatorio 3, IT-35122, Padova, Italy\label{inst3}
             \and Space Science and Astrobiology Division, NASA Ames Research Center, MS 245-3, Moffett Field, CA 94035, USA\label{inst4}
             \and
             INAF - Osservatorio Astrofisico di Arcetri, Largo E. Fermi 5, 50125, Firenze, Italy\label{inst5}
             \and INAF - Osservatorio Astronomico di Palermo, Piazza del Parlamento 1, I-90134, Palermo, Italy\label{inst6}
             \and INAF - Osservatorio Astronomico di Trieste, Via Giambattista Tiepolo, 11, I-34131, Trieste (TS), Italy\label{inst7}
             \and
             ESO - European Southern Observatory, Alonso de Córdova 3107, Casilla 19, Santiago, 19001, Chile\label{inst8}
             \and INAF – Osservatorio Astronomico di Roma, Via Frascati 33, 00078, Monte Porzio Catone (Roma), Italy\label{inst9}
             \and
             Center for Astrophysics \textbar \ Harvard \& Smithsonian, 60 Garden Street, Cambridge, MA 02138, USA\label{inst10}
             \and 
             INAF -- Osservatorio Astrofisico di Torino, Via Osservatorio 20, IT-10025, Pino Torinese, Italy\label{inst11}
             \and
             Komaba Institute for Science, The University of Tokyo, 3-8-1 Komaba, Meguro, Tokyo 153-8902, Japan\label{inst12}
             \and
              NSF NOIRLab, 950 N. Cherry Ave., Tucson, AZ 85719, USA\label{inst13}
              \and Dipartimento di Fisica, Universit\'a degli Studi di Torino, via Pietro Giuria 1, I-10125, Torino, Italy\label{inst14}
              \and
             Department of Astronomy, California Institute of Technology, Pasadena, CA 91125, USA\label{inst15}
             \and INAF - Osservatorio Astrofisico di Catania, Oss. Astr. Catania, via S. Sofia 78, 95123 Catania Italy\label{inst16}
             \and 
              Department of Physics and Astronomy, Vanderbilt University, Nashville, TN 37235, USA\label{inst17}
              \and
              Department of Physics \& Astronomy, University of California Los Angeles, Los Angeles, CA 90095, USA\label{inst18}
              \and
             Department of Astronomy, University of California, Berkeley, Berkeley, CA 94720, USA\label{inst19}
             \and
             Instituto de Astrofisica de Canarias (IAC), 38205 La Laguna, Tenerife, Spain\label{inst20}
             \and Fundación Galileo Galilei – INAF, Rambla José Ana Fernandez Pérez 7, 38712 Breña Baja, TF, Spain\label{inst21}
            \and
             Graduate School of Social Data Science, Hitotsubashi University, 2-1 Naka, Kunitachi, Tokyo 186-8601, Japan\label{inst22}
            \and
             Department of Physical Sciences, Ritsumeikan University, Kusatsu, Shiga 525-8577, Japan\label{inst23}
             \and
             Lund Observatory, Division of Astrophysics, Department of Physics, Lund University, Box 118, 22100 Lund, Sweden\label{inst24}
             \and
             Departamento de Astrof\'isica, Universidad de La Laguna (ULL), E-38206 La Laguna, Tenerife, Spain\label{inst25}
             \and
             Dipartimento di Fisica, Università di Roma Tor Vergata, Via della Ricerca Scientifica 1, 00133, Roma, Italy\label{inst26}
             \and
             Max Planck Institute for Astronomy, Königstuhl 17, 69117, Heidelberg, Germany\label{inst27}
             \and
             Astrobiology Center, 2-21-1 Osawa, Mitaka, Tokyo 181-8588, Japan\label{inst28}
             \and
             National Astronomical Observatory of Japan, 2-21-1 Osawa, Mitaka, Tokyo 181-8588, Japan\label{inst29}
             \and
             Department of Astronomy, University of Maryland, College Park, College Park, MD 20742, USA\label{inst30}
             \and
             Department of Physics and Kavli Institute for Astrophysics and Space Research, Massachusetts Institute of Technology, Cambridge, MA 02139, USA\label{inst31}
             \and
             Kotizarovci Observatory, Sarsoni 90, 51216 Viskovo, Croatia\label{inst32}
             \and
             Hazelwood Observatory, Australia\label{inst33}
             \and
             Planetary Discoveries, Valencia, CA 91354, USA\label{inst34}
             \and
             Società Astronomica Lunae, Castelnuovo Magra, Italy\label{inst35}
             }

    \date{Compiled: \today}
 
  \abstract
   {The present-day architecture of planetary systems contains information about their formation and migration histories. The origin of hot Jupiters (HJs; P $\lesssim$ 10 d, $R_{\rm p} > 8 R_\oplus$) has long been a matter of debate. While most of them are found to be `lonely', there is a rare population of HJs hosting small inner-orbit companions (eight known as of May 2026). Their peculiar architecture suggests a gentle disc-migration mechanism. In this study, we present the discovery and characterisation of the multi-planet system TOI-1533, comprising an inner sub-Neptune (TOI-1533~b, $P_{\rm orb} = 3.63$ d, $R_{\rm p} = 3.15~R_\oplus$) and an outer hot giant planet (TOI-1533~c, $P_{\rm orb} = 8.06$ d, $R_{\rm p} > 7.5~R_\oplus$) with substantial H/He by mass ($\rho_{\rm p} < 0.48$ g cm$^{-3}$), both transiting an active K-dwarf star ($T_{\rm eff} \approx$ 5130 K; $V$ (mag) $\approx$ 11). Our joint modelling of stellar activity and planetary signals from radial velocities (HARPS-N) and transits (TESS) allowed us to detect the planetary Keplerian signals (approximately $10~\sigma$) and isolate the stellar modulation. The inclusion of simultaneous photometry in the multi-dimensional Gaussian processes formalism was a fundamental addition to the spectroscopic activity indicators, enabling the disentanglement of stellar activity from planetary signals. The mass ratio of the two confirmed planets ($M_{\rm b} / M_{\rm c}$ about 0.8), together with the super-Neptune mass of the large outer companion ($M_{\rm c} \approx 40 M_\oplus$), makes this system unusual compared to the other known HJs with low-mass inner companions. }


    \titlerunning{TOI-1533: a compact system hosting a super-Neptune-mass pair with disparate radii}

   \maketitle

\section{Introduction}
The origin of gas giants on close-in orbits (P $\lesssim$ 10 d, $R_{\rm p} > 8 R_\oplus$), known as hot Jupiters (HJs), is a long-standing question in planet formation and evolution theories. Although many scenarios have been proposed to place HJs in their current orbits (e.g. disc migration, high-eccentricity migration), none can satisfy all the observational constraints \citep{2018ARA&A..56..175D}. The answer seems hidden in the architectures of their systems, which remain only partially understood. One clue is that HJs are typically `lonely', and their companions, if present, are often massive and on wide orbits \citep[$P > 200$ d,][]{2016ApJ...825...62S}. 

The lack of small, low-mass, and short-period companions initially suggested that HJs form beyond the snow line and then migrate inwards via high-eccentricity migration \citep[][]{2003ApJ...589..605W}. In this scenario, the eccentricity of a gas giant is excited by close encounters with other giant planets \citep[e.g.][]{rasioford1996} or by Kozai's interaction with a distant stellar companion \citep{2001ApJ...562.1012E}. If the periastron distance is small enough, the orbit circularises and shrinks due to star-planet tidal interaction. Such a dynamical instability is destructive for low-mass bodies in the path of the migrating giant \citep{Mustill2015} and would inevitably leave HJs in `loneliness'. However, among the 500 plus known HJs, a growing number of exceptions host low-mass inner companions. These are WASP-47 \citep{2015ApJ...812L..18B}, Kepler-730 \citep{2019ApJ...870L..17C}, TOI-1130 \citep{Huang_2020}, WASP-132 \citep{2022AJ....164...13H}, TOI-2000 \citep{2023MNRAS.524.1113S}, TOI-5398 \citep{2022MNRAS.516.4432M, 2024AA...682A.129M}, TOI-1408 \citep{2024ApJ...971L..28K}, and TOI-5143 \citep{2026AJ....171..359Q}. Their peculiar architecture suggests an alternative, gentler migration mechanism.

Under disc migration, tidal interactions with the protoplanetary disc move the initially cold giant inward from its birthplace \citep{1996Natur.380..606L}, sweeping material along its mean-motion resonances to form inner planets \citep{2006Sci...313.1413R}. Measuring the occurrence rate of these inner companions can help determine the fraction of HJs that formed via disc migration versus high-eccentricity migration \citep{2023MNRAS.524.1113S}. Precise masses and orbital parameters are needed to study the formation and migration scenarios of these maverick systems \citep[e.g.][]{2022NatAs...6..736S}. At present, however, the sample of systems with companions is too small and relatively uncharacterised to answer this question.

In this work, we present the discovery and characterisation of a new system of this kind, TOI-1533. All the data and observations collected for this purpose are described in Sect. \ref{sec:obs}, while we describe TOI-1533's stellar properties in Sect. \ref{sec:stellar}. We present our joint modelling of the photometric and spectroscopic data used to confirm both planets in Sect. \ref{sec:analysis}. In Sect. \ref{sec:discussion} we discuss our results and present the specific and global characteristics of this new multi-planet system comprising a super-Neptune-mass pair with disparate radii. Concluding remarks are reported in Sect. \ref{sec:conclusions}.

\section{Observations and data reduction}
\label{sec:obs}

\subsection{TESS photometry}
The planetary system presented in this work was first detected in the Transiting Exoplanet Survey Satellite (\textit{TESS}) photometry (TOI-1533.01, $P \approx 3.6$ d; TOI-1533.02, $P \approx 8.1$ d). In particular, \textit{TESS} observed TOI-1533 (TIC 345143460) with a 30~min cadence in Sectors 17 and 24 and with a 2~min cadence in Sectors 57, 77, 78, 84, and 85. We extracted \textit{TESS} light curves following the \texttt{PATHOS} approach \citep[see][for a thorough description]{2020MNRAS.495.4924N}, which also allowed us to minimise neighbour flux contamination \citep[see e.g.][]{2022A&A...664A.163N, 2024AA...682A.129M}. 

\subsection{ASAS-SN photometry} 
We inspected almost 6 years of archival data from ASAS-SN \citep{2014ApJ...788...48S,2017PASP..129j4502K}, with observations spanning from June 2019 to November 2024, to estimate the stellar rotation period (see Sect. \ref{sec:rot_per}). The data for TOI-1533 were taken in the Sloan $g-$band and consist of images with a resolution of 8 arcsec/pixel, $\sim$15$\arcsec$ full width at half-maximum (FWHM) point spread function.

\subsection{Planet vetting and validation} 
\label{sec:groundtransits}

This subsection presents the data aimed at demonstrating that both candidates, TOI-1533.01, TOI-1533.02, in the system are high-confidence planets (from now on TOI-1533~b, TOI-1533~c). With multiple uncontaminated ground-based observations as part of the TESS Follow-up Observing Program \citep[Sub Group 1, TFOP][]{2019AAS...23314005C}\footnote{https://tess.mit.edu/followup}, we detected both planets and excluded deep false-positive eclipses. We used the {\tt TESS Transit Finder}, a customised version of the {\tt Tapir} software package \citep{Jensen:2013}, to schedule our follow-up transit observations. Moreover, recon spectroscopic analyses ruled out the presence of a spectroscopic binary. We present the data and our analysis later in Sect. \ref{sec:analysis}.

\subsubsection{LCOGT}
\label{section:lcogt}
We observed full transit windows of TOI-1533~b in the Pan-STARRS $z_s$ band on UT 2023 December 3; 2024 August 4; 2024 August 11 from the Las Cumbres Observatory Global Telescope \citep[LCOGT;][]{Brown:2013} 1.0\,m network nodes at the Teide Observatory on the island of Tenerife (Teide) and the McDonald Observatory near Fort Davis, Texas, United States (McD), respectively. We observed a full transit window of TOI-1533~c in the Sloan $i'$ band on UT 2023 October 7 from the LCOGT 0.35\,m network node at the Haleakala Observatory on Maui, Hawai'i (Hal). We also observed full transit windows of TOI-1533~c on UT 2024 October 12 and 2025 November 3 from the LCOGT 2.0\,m Faulkes Telescope North at the Haleakala Observatory. The Faulkes Telescope North is equipped with the MuSCAT3 multi-band imager \citep{Narita:2020}. All images were calibrated with {\tt BANZAI} \citep{McCully:2018}, and photometric data were extracted using {\tt AstroImageJ} \citep{Collins:2017}. Circular apertures with 4 -- 8$\arcsec$ radii were used to extract the differential photometry. All apertures excluded flux from the nearest Gaia DR3 star (Gaia DR3 1998672965462506496), which is 13" southwest of TOI-1533, confirming both transit signals on the target. MuSCAT3 observations across the Sloan $g'$, $r'$, $i'$, and $z_s$ bands show no indication of transit depth chromaticity in TOI-1533~c transits. All light curves are included in the joint modelling described in Sect. \ref{sec:modelling}.

\subsubsection{MuSCAT2}
\label{section:MuSCAT2}
We observed two full transits of TOI-1533~b on 2023 November 11 and 2024 September 23 UT under variable weather conditions, with intermittent cloud passages, and one full transit of TOI-1533~c on 2023 December 26 under clear conditions. The observations were obtained in the $g'$, $r'$, $i'$, and $z_s$ bands using the MuSCAT2 \citep{Narita2019} instrument mounted on the 1.52\,m TCS telescope at Teide Observatory in the Canary Islands, Spain. Exposure times ranged from 5 to 15\,s depending on the observing conditions. A slight telescope defocus was applied to improve photometric precision while maintaining the resolution of a nearby star located $13''$ from the target.

Data reduction and aperture photometry were performed using the MuSCAT2 pipeline,\footnote{https://github.com/hpparvi/MuSCAT2\_transit\_pipeline} described in \citet{Parviainen2020}. The pipeline evaluates multiple combinations of comparison stars and aperture radii. Optimal light curves were selected through a global optimisation procedure that fits a model including the five brightest comparison stars within the field of view and uncontaminated photometric apertures. We could only marginally detect the TOI-1533~b transit due to unfavourable weather, whereas we detected an on-time and achromatic transit of TOI-1533~c with high significance.

\subsubsection{Adaptive optics imaging}
We observed TOI-1533 on UT 2020 December 1 using the ShARCS camera on the Shane 3-meter telescope at the Lick Observatory \citep{2012SPIE.8447E..3GK, 2014SPIE.9148E..05G, 2014SPIE.9148E..3AM}. Observations were taken with the Shane adaptive optics system in natural guide star mode to search for nearby unresolved stellar companions. We collected a single sequence of observations using a $Ks$ filter ($\lambda_0 = 2.150$ $\mu$m, $\Delta \lambda = 0.320$ $\mu$m). We reduced the data using the publicly available \texttt{SImMER} pipeline \citep{2020AJ....160..287S}.\footnote{https://github.com/arjunsavel/SImMER} Our reduced images and corresponding contrast curves are shown in Fig. \ref{fig:sharcs}. Our observations achieved a contrast of 3.0 at 0.5$\arcsec$ and 4.5 at 1.0$\arcsec$. We found no nearby stellar companions within our detection limits.

\subsubsection{Speckle imaging}
We observed TOI-1533 on UT 2022 September 17 using the NN-EXPLORE Exoplanet Stellar Speckle Imager \citep[NESSI;][]{2018PASP..130e4502S}, a speckle imager at the WIYN 3.5~m telescope on Kitt Peak. The data consist of 7000 40~ms frames in two filters with central wavelengths $\lambda_c = 562$ and 832~nm. NESSI's field of view was confined to a $256\times256$ pixel sub-array readout ($4.6\arcsec\times4.6\arcsec$).  However, our speckle measurements were further restricted to an outer radius of $1.2\arcsec$ from the target star. A similar set of 1000 speckle frames was taken of a nearby single star in order to calibrate the point spread function of the TOI-1533 data.

These speckle data were reduced following \cite{2011AJ....142...19H}. We obtained a contrast curve from each reconstructed image of the field by measuring fluctuations in the noise-like background level as a function of separation from TOI-1533 (Fig. \ref{fig:nessi}). These contrast curves establish the upper limits on the relative brightness of any undetected point source in proximity to the target star. No companion sources were detected for TOI-1533 in the NESSI data.

\subsubsection{Reconnaissance spectroscopy}
TOI-1533 was observed with the HIRES \citep[HIgh Resolution Echelle Spectrometer][]{2010ApJ...721.1467H} spectrograph mounted on the Keck telescope as part of the California Planet Survey between December 2019 and June 2020. These observations confirmed the absence of any sign of a spectroscopic binary down to a flux ratio of 1\%, and they demonstrated that the star is quite active, with an S-index value from the highest S/N spectrum of $0.432 \pm 0.05$ \citep{2024ApJS..274...35I} and $\log{R'_{\rm HK}} = -4.57 \pm 0.06$, using $(B-V)_0$ from Table \ref{tab:star_param}. TRES echelle spectrograph \citep{2007RMxAC..28..129S} observations spread out over two years showed a downward drift of about 200 m s $^{-1}$, but they also exclude any sign of a false positive scenario. 

\subsection{HARPS-N spectroscopic follow-up}
\label{sec:harpn}

We observed TOI-1533 with the High Accuracy Radial velocity Planet Searcher for the Northern hemisphere \citep[HARPS-N;][]{2012SPIE.8446E..1VC} spectrograph at TNG between October and December 2024, obtaining a total of 17 spectra with 1200 s of exposure time with an average S/N of 44 at 5500~\AA~and spanning a time window of 82 days. It is worth noting that \textit{TESS} sectors 84 and 85 overlap with the spectroscopic observations. The spectra, which cover the wavelength range 383 -- 693 nm and have a resolving power of $\approx$ 115~000, were collected under the open-time proposal A50TAC\_30 (PI: G. Mantovan). We agreed to share the awarded time with other Global Architecture of Planetary Systems \citep[GAPS,][]{2013A&A...554A..28C} programmes requesting radial velocity (RV) monitoring to ensure the optimisation and coordination of the observations. 

The HARPS-N RV data reduction was carried out in a way similar to that presented in several previous GAPS papers. In particular, we used the HARPS-N DRS v3.2.0, derived from the ESPRESSO DRS \citep{2021A&A...648A.103D}, and computed the RVs using the cross-correlation function (CCF) method \citep{2002Msngr.110....9P} with a G9 mask. To better distinguish the activity contribution from the Keplerian signals in the RV dataset, we additionally extracted multiple activity indicators. Specifically, from the HARPS-N DRS we extracted the value of the CCF bisector span, the full width at half-maximum (FWHM), and its equivalent width \citep[$W_{\rm CCF}$][]{2017MNRAS.469.3965M,2019MNRAS.487.1082C}. The chromospheric $\log R'_{\rm HK}$ index was instead obtained using a method from \cite{2011arXiv1107.5325L} available on the YABI workflow interface \citep{Hunter2012} installed at the IA2 data centre.\footnote{\url{https://www.ia2.inaf.it/}}

\subsection{NEID spectroscopic follow-up}
\label{sec:neid}
We observed TOI-1533 with the NEID spectrograph \citep{NEID_Schwab2016,NEID_Halverson2016} on the WIYN 3.5m telescope at the Kitt Peak National Observatory (KPNO). We obtained seven spectra between 2025 July 10 and 2025 October 31 in high-resolution mode ($R \sim 110{,}000$) with exposure times of 480s, attaining S/N of 25--35 at 5500~\AA. The data were reduced using v1.4.2 of the standard NEID Data Reduction Pipeline (DRP),\footnote{\url{https://neid.ipac.caltech.edu/docs/NEID-DRP/}} with precise RVs extracted by cross-correlating the spectrum with a weighted stellar line mask for spectral type K2 \citep{Baranne1996,Pepe2002}. Stellar activity indicators were also extracted from each spectrum using the standard DRP algorithms.

\section{Stellar parameters}
\label{sec:stellar}

We determined the stellar parameters of TOI-1533 using photometric, astrometric, spectroscopic, and further ancillary data. The whole procedure was started by producing a co-added spectrum using all the HARPS-N spectra, with an average S/N of $\sim$ 230 per extracted pixel at around 6000 Å.

\subsection{Atmospheric parameters and iron abundance}
\label{sec:atm_param}

We analysed the co-added HARPS-N spectrum to derive the effective temperature $T_{\rm eff}$, the surface gravity $\log g$, the iron abundance $\rm [Fe/H]$, and the microturbulence velocity $\xi$ following the same methodology as in \cite{2025basilicata}. We employed the standard equivalent width method with \ion{Fe}{I} and \ion{Fe}{II} lines from the \cite{2022A&A...664A.161B} list. Briefly, we measured equivalent widths with the automatic tool \texttt{ARES} (v2, \citealt{2015sousa}) and by using the \texttt{qoyllur--quipu} software (q2\footnote{\url{https://github.com/astroChasqui/q2}}, \citealt{2014ramirez}). We adopted the 1D local thermodynamic equilibrium model atmospheres with new opacities from the \cite{2003IAUS..210P.A20C} grid. The final values of the spectroscopic stellar parameters can be found in Table \ref{tab:star_param}, suggesting a K1-type dwarf \citep[from][version 2022.04.16]{2013ApJS..208....9P} with super-solar chemical composition. Our final value of $T_{\rm eff, spec}$ agrees perfectly with the estimates obtained with the \textit{colte} tool \citep{2021colte}, using various colour indexes from 2MASS \citep{2003tmc..book.....C} and \textit{Gaia}-DR3 \citep{2023A&A...674A...1G} photometry. The mean photometric value of $T_{\rm eff, phot}$ is $5127\pm73$\,K, spanning from a minimum in $(G-G_{\rm BP})$ of $5023\pm54$\,K and a maximum in $(G_{\rm RP} - H)$ of $5238\pm72$\,K. The spectroscopic $\log g$ is also in perfect agreement with the value reported in the \textit{Gaia} catalogue of $4.53 \pm 0.03$. Finally, the microturbulence velocity we obtained is comparable within the uncertainties to the value of $0.80\pm0.05$\,km\,s$^{-1}$ measured via the empirical relation of \citep{2016dutra}.

\subsection{Infrared flux method and isochrones}
\label{sec:irfm}
We derived the stellar radius and mass of TOI-1533 from its atmospheric parameters ($T_{\rm eff}$, $\log{g}$) using a custom-built code that combines the infrared flux method \citep[IRFM,][]{1977MNRAS.180..177B, 2017AJ....153..136S} and isochrone fitting \citep[MCMCI code,][]{2020A&A...635A...6B}. The process is fully described, for example, in \cite{2026A&A...709A.265B} and \cite{2026arXiv260504149M}. The resulting values are reported in Table~\ref{tab:star_param}. Stellar atmospheric models were taken from the ATLAS catalogues \citep{2003IAUS..210P.A20C}.

\subsection{Chromospheric activity}
\label{sec:chrom}
The lower chromosphere Ca II H\&K emission was measured on HARPS-N spectra using YABI (see Sect. \ref{sec:harpn}). The average value of the S-index calibrated on the Mt. Wilson scale \citep{1995ApJ...438..269B} is 0.45, which corresponds to $\log R'_{\rm HK}$ = $-$4.55 $\pm$ 0.03. For the $\log R'_{\rm HK}$ determination, we adopted the colour index $(B-V)_0$ = 0.92 mag, which was derived from the APASS DR10 \citep{2016yCat.2336....0H} observed values.

\subsection{Rotation period}
\label{sec:rot_per}
The wide dataset we collected allowed us to perform a precise estimate of the stellar rotation period, which is a key parameter for age determination and star--planet system orientation. First, we inspected the ASAS-SN photometry and ran the Generalised Lomb-Scargle (GLS) periodogram algorithm \citep{2009A&A...496..577Z}, which revealed the most powerful peak with a period of about 20.8 days. 

We then estimated the rotation period in our fully Bayesian analysis (see Sect. \ref{sec:analysis}) by leaving it free to vary and modelling the activity in our \textit{TESS}-corrected light curves (which preserve stellar variability), RV, FWHM, and $\log R'_{\rm HK}$ series. To support our findings, we used the GLS periodogram analysis to examine the two spectroscopic activity indicators and the RVs. All the periodograms show a dominant peak close to 20-22 days. Most \textit{TESS} sectors also show the highest peak around 21 days, with occasional signals close to 10 or 16 days. The latter periodicities could occur, for example, if there were two similarly sized spots in opposite hemispheres.

Based on these considerations, we adopted the $P_{\rm rot}$ value from our full analysis (see Table \ref{table:model-lcrv}) as the reference for this work, while we interpreted the 10-day signal as the first harmonic. As we detail below, this interpretation is further supported by multiple diagnostics. The expected rotation period from the $\log R'_{\rm HK}$ value is $19.9$~d using the \cite{2008ApJ...687.1264M} calibration. This further supports the adopted period as the true one.

\begin{table}[!t]
\tiny
   \caption[]{Stellar properties of TOI-1533.}
     \label{tab:star_param}
     \centering
       \begin{tabular}{lcc}
         \hline
         \noalign{\smallskip}
         Parameter   &  \object{TOI-1533} & Reference  \\
         \noalign{\smallskip}
         \hline
         \noalign{\smallskip}
$\alpha$ (J2000)                &   $+$23:40:59.21      & {\it Gaia} DR3  \\
$\delta$ (J2000)                &   $+$57:29:06.87   & {\it Gaia} DR3  \\
$\mu_{\alpha}$ (mas yr$^{-1}$)  &    $+$19.78 $\pm$ 0.01  & {\it Gaia} DR3  \\
$\mu_{\delta}$ (mas yr$^{-1}$)  &    $-$34.47 $\pm$ 0.01  & {\it Gaia} DR3  \\
RV     (km s$^{-1}$)            &    $-$23.73 $\pm$ 0.30   & {\it Gaia} DR3  \\
$\pi$  (mas)                    &     10.00 $\pm$ 0.01   & {\it Gaia} DR3  \\
RUWE                     &     1.1   & {\it Gaia} DR3  \\
\noalign{\medskip}
$V$ (mag)                 &  10.98 $\pm$  0.04  &   \cite{2016yCat.2336....0H}   \\
$B-V$ (mag)             &  0.92 $\pm$  0.04     &   \cite{2016yCat.2336....0H}  \\
$G$ (mag)               &  10.6843 $\pm$ 0.0005 &   {\it Gaia} DR3  \\
$G_{\rm BP}$ (mag)               &  11.156 $\pm$ 0.002 &   {\it Gaia} DR3  \\
$G_{\rm RP}$ (mag)               &  10.058 $\pm$ 0.001 &   {\it Gaia} DR3  \\
$G_{\rm BP}-G_{\rm RP}$ (mag)   &        1.098 $\pm$ 0.002      &   {\it Gaia} DR3  \\
$J$ (mag)    &   9.34 $\pm$ 0.03 & 2MASS  \\
$H$ (mag)    &   8.96 $\pm$ 0.03 & 2MASS  \\
$K$ (mag)    &   8.83 $\pm$ 0.02 & 2MASS  \\
\noalign{\medskip}
$T_{\rm eff, spec}$ (K)        &   5146 $\pm$ 81  & This paper (Sect. \ref{sec:atm_param}) \\  
$\log g$                 &   4.50 $\pm$ 0.15   & This paper (Sect. \ref{sec:atm_param}) \\ 
$\xi$                 &   1.09 $\pm$ 0.20   & This paper (Sect. \ref{sec:atm_param}) \\ 
${\rm [Fe/H]}$ (dex)     &   +0.24 $\pm$ 0.06   & This paper (Sect. \ref{sec:atm_param}) \\ 
$\log R'_{\rm HK}$    &     $-$4.55 $\pm$ 0.03 &  This paper (Sect. \ref{sec:chrom}) \\  
$v\sin{i_{\star}}$ (km s$^{-1}$)      &   2.5 $\pm$ 0.5  & This paper (Sect. \ref{sec:lithium}) \\  
$P_{\rm rot}$ (d)  &    20.02$^{+0.37}_{-0.33}$ & This paper (Sect. \ref{sec:analysis}) \\
$EW_{\rm Li}$ (m\AA)     &  1.9 $\pm$ 0.3 &  This paper (Sect. \ref{sec:lithium})  \\
\noalign{\medskip}
Luminosity ($L_{\odot}$) &    0.45 $\pm$ 0.03 & This paper  (Sect. \ref{sec:irfm}) \\
Radius ($R_{\odot}$)     &    0.849 $\pm$ 0.004 & This paper  (Sect. \ref{sec:irfm}) \\
Mass ($M_{\odot}$)       &    0.88 $\pm$ 0.04 & This paper  (Sect. \ref{sec:irfm}) \\
Density ($\rho_{\odot}$)       &    1.44 $\pm$ 0.07 & This paper  (Sect. \ref{sec:irfm}) \\
E(B-V)  &    0.041 $\pm$ 0.003 & This paper  (Sect. \ref{sec:irfm}) \\
Age  (Gyr)               &    6.9$^{+1.8}_{-3.4}$ & This paper  (Sect. \ref{sec:irfm})  \\
Distance  (pc)           &   99.7$^{+0.4}_{-0.3}$ & {\it Gaia} DR3 \rule{0pt}{2.5ex} \rule[-1ex]{0pt}{0pt} \\
         \noalign{\smallskip}
         \hline
      \end{tabular}
\end{table}

\subsection{Lithium abundance and projected rotational velocity}
\label{sec:lithium}
As an additional age estimate, we examined the possible presence of the lithium absorption line at $\lambda$~6707 \AA. The line was indeed detected, and we measured an equivalent width of 1.9 $\pm$ 0.3 m\AA. Combining this value with the effective temperature derived above and applying the EAGLES \citep[Empirical AGes from lithium Equivalent widthS;][]{2023MNRAS.523..802J} code, we derived a lower limit of 800 Myr. The projected rotational velocity of TOI-1533 was determined using the calibration between the FWHM of the CCF and $v\sin i_\star$ developed by \cite{Rainer2023}, yielding a value of $v\sin i_\star$ = 2.5 $\pm$ 0.5 km s$^{-1}$.

\section{Analysis}
\label{sec:analysis}
\label{phot}
\subsection{Joint times series Bayesian analysis}
\label{sec:modelling}

To assess the planetary properties of TOI-1533 b and TOI-1533 c, we modelled all the \textit{TESS}-corrected light curves and ground-based ones simultaneously with the HARPS-N spectroscopic time series. We did so by using \texttt{PyORBIT}\footnote{\url{https://pyorbit.readthedocs.io/}} \citep{2016A&A...588A.118M, 2018AJ....155..107M}, a Python package that models planetary transits and radial velocity signals while considering stellar activity effects. We tried various approaches to model the activity using Gaussian processes \citep[GPs,][]{rasmussen2006gaussian, 2014MNRAS.443.2517H}.

We modelled the planetary transits, the stellar activity, and the Keplerian signals within the RV series by simultaneously fitting the time of inferior conjunction ($T_0$); the planetary-to-star radius ratio ($R_{\rm p}/R_{\star}$); the impact parameter ($b$); the orbital period ($P_{\rm orb}$); the RV semi-amplitude ($K$); the stellar density ($\rho_\star$); the quadratic limb-darkening (LD) coefficients $u_1$ and $u_2$, adopting the LD parameterisation introduced by \cite{2013MNRAS.435.2152K}; the systemic RV (offset); and a jitter term added in quadrature to measurement errors for each photometric and RV dataset. This accounts for effects not modelled (e.g. short-term stellar activity) or potential underestimation of existing error bars. We fitted the orbital periods and semi-amplitudes in linear space. We assumed a circular orbit for both planets, while we imposed Gaussian priors on $u_1$ and $u_2$. These coefficients were estimated using \texttt{PyLDTk}\footnote{\url{https://github.com/hpparvi/ldtk}} \citep{2013A&A...553A...6H,2015MNRAS.453.3821P}, and we took into account the specific filters used during the observations and added $0.1$ in quadrature to their Gaussian errors to account for the known model underestimation. The complete list of priors can be found in Table \ref{table:model-lcrv}. 

We modelled stellar activity in the RV, FWHM, $\log R'_{\rm HK}$, and \textit{TESS} sector 84 time series using a multidimensional GP, while we used a unidimensional one for all other \textit{TESS} photometric time series. We used an exponential-sine periodic (ESP) kernel as defined in \citet{2020A&A...638A..95D, 2022A&A...659A.182D} for the multidimensional GP, and a rotation kernel as defined in \citet{celerite1, celerite2} for the unidimensional ones. A unique stellar rotation period ($P_{\rm rot}$) was shared between the multidimensional GP and the \texttt{celerite2} GPs. The other GP hyperparameters, such as the characteristic decay timescale ($P_{\rm dec}$) and coherence scale ($\omega_{\rm cs}$), are independent between the photometric and spectroscopic datasets. 

\begin{figure}
   \centering
   \includegraphics[width=0.7\hsize]%
   {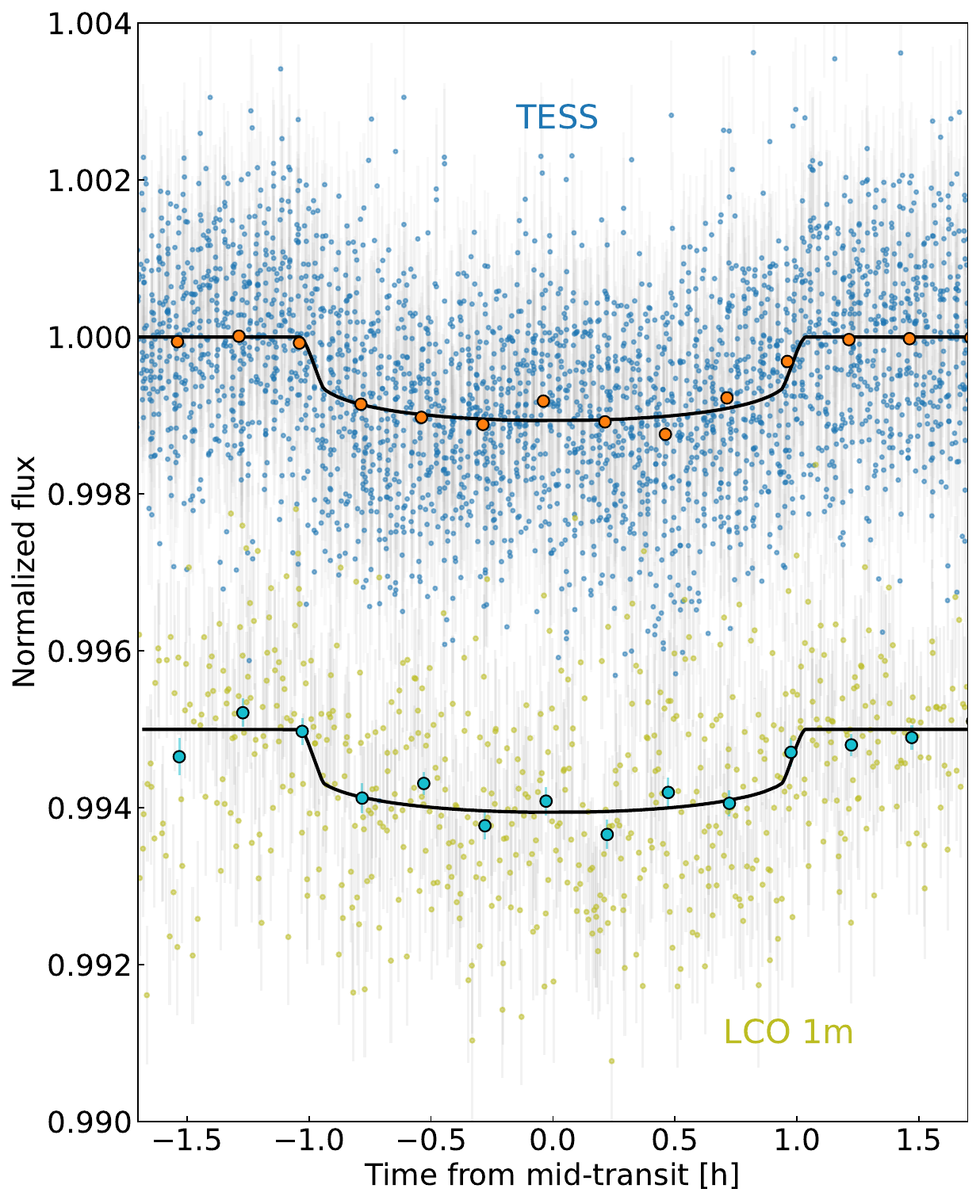}
   \caption{Photometric modelling of the TOI-1533 b planetary signal. \textit{TESS} phase-folded transits of TOI-1533 b are shown after normalisation together with the transit model (black line). Ground-based light curves are included as well.}
   \label{fig:1533b_lc}
\end{figure}

\begin{figure}
   \centering
   \includegraphics[width=0.7\hsize]%
   {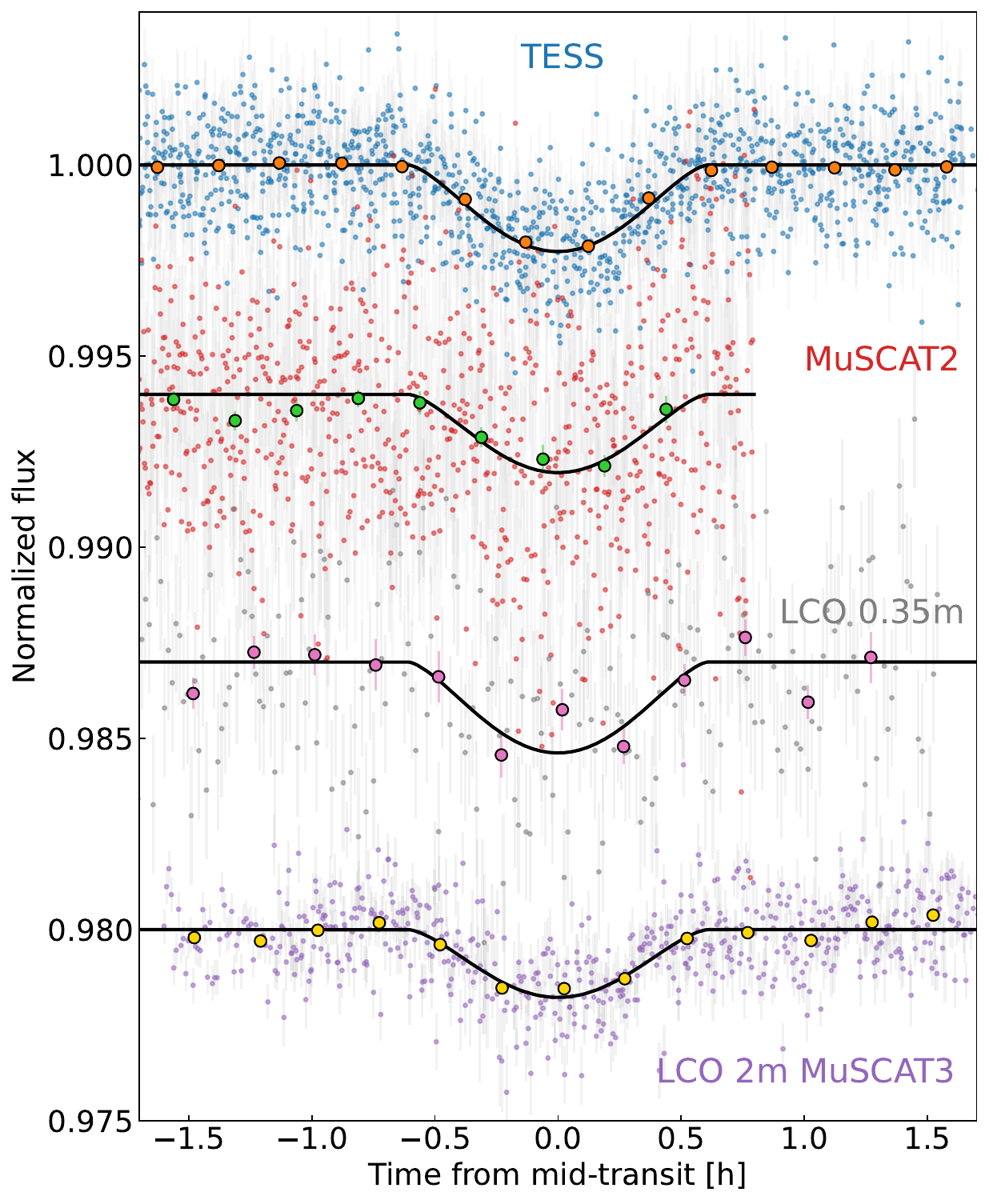}
   \caption{Same as Figure \ref{fig:1533b_lc} but for planet c. The MuSCAT2 light curve is in the Sloan $i'$ filter, while the MuSCAT3 ones are in the Sloan $g'$ filter.}
   \label{fig:1533c_lc}
\end{figure}

\begin{figure}
   \centering
   \includegraphics[width=0.495\hsize]%
   {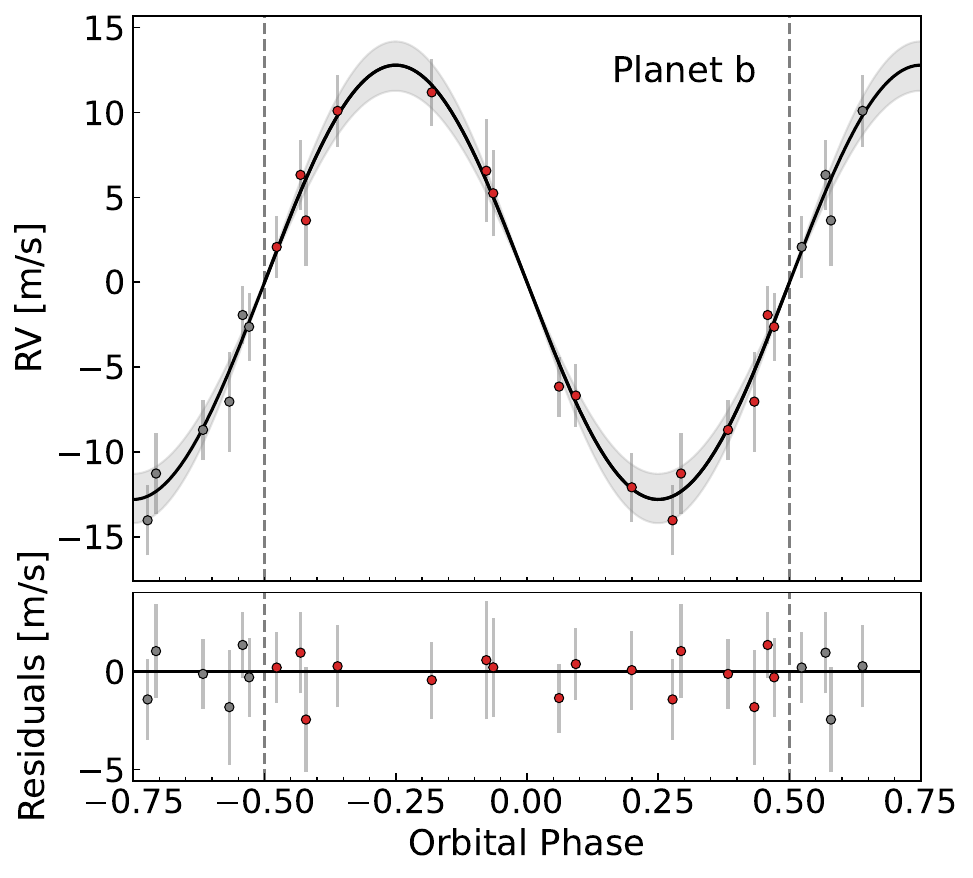}
   \includegraphics[width=0.495\hsize]%
   {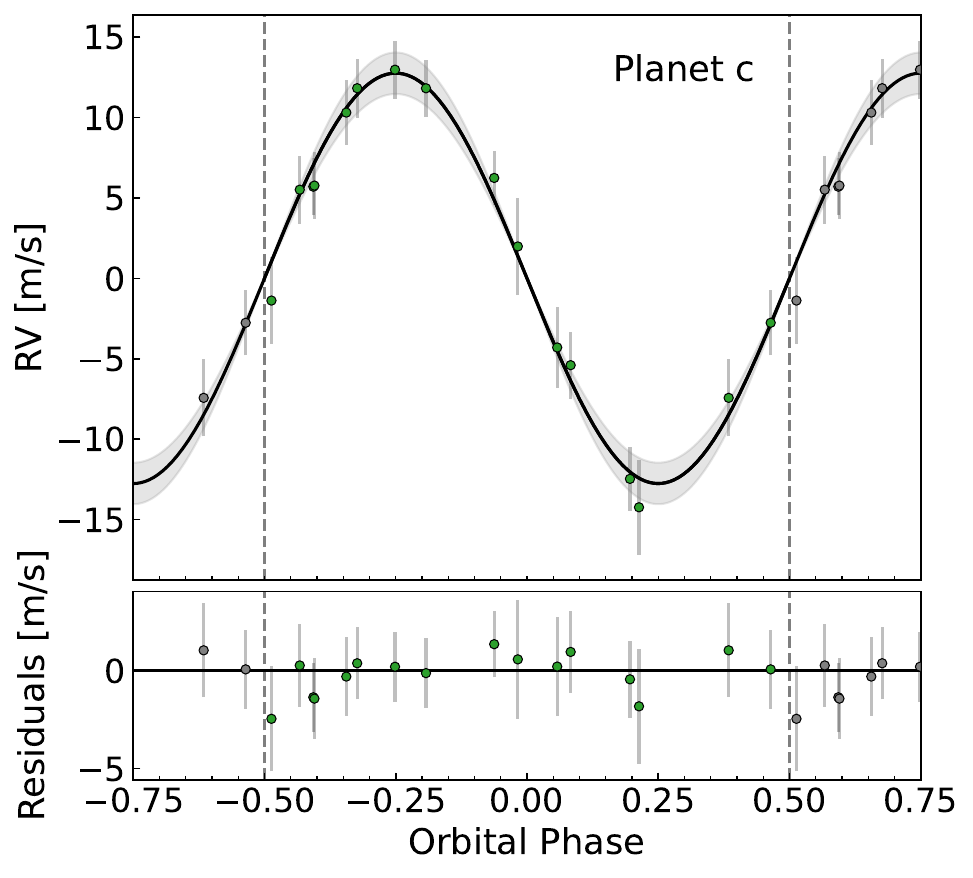}
   \caption{Phase-folded RV curves of TOI-1533 b and c. The shaded area shows the uncertainties ($\pm 1\sigma$) of the RV model. The residuals of the fit are shown in the \textit{bottom} panel. }
   \label{fig:1533rv}
\end{figure}

We performed global optimisation of the parameters by running \texttt{PyDE} \citep{1997JGOpt..11..341S, 2016zndo.....45602P} for 50~000 generations and then conducted a Bayesian analysis using \textsc{emcee} \cite{2013PASP..125..306F} for 100~000 steps. We used $4\times n_{\rm dim}$ walkers, with $n_{\rm dim}$ as the model dimensionality, and we discarded the first 25~000 steps as burn-in. We applied a thinning factor of 100 to mitigate chain auto-correlation. Figures \ref{fig:1533b_lc}, \ref{fig:1533c_lc},\ref{fig:1533rv}, and\ref{fig:rv_full} and Tables \ref{table:model-lcrv} and \ref{table:model-full} show the results of the modelling.

\begin{table*}
\centering
{\tiny
\caption{Priors and confidence intervals of the main parameters from spectroscopic plus photometric modelling.}             
\label{table:model-lcrv}    
\addtolength{\tabcolsep}{-0.15em}
\begin{tabular}{l c c c c c}     
\hline\hline 
\multicolumn{2}{c}{Planetary parameters} & \multicolumn{2}{c}{TOI-1533 b} & \multicolumn{2}{c}{TOI-1533 c}\rule{0pt}{2.3ex} \rule[-1ex]{0pt}{0pt}\\ 
\hline
Parameter & Unit & Prior & Value & Prior & Value\rule{0pt}{2.3ex} \rule[-1ex]{0pt}{0pt}\\ 
\hline
   Orbital period ($P_{\rm orb}$) & days & $\mathcal{U}$(3.5, 3.7) & 3.6457974 $\pm$ 0.0000034  & $\mathcal{U}$(8.0, 8.1) & 8.0637926 $\pm$ 0.0000052 \rule{0pt}{2.2ex} \rule[-1ex]{0pt}{0pt}\\
   Central time of transit ($T_{\rm 0}$) & BTJD & $\mathcal{U}$(2877.6, 2877.8) & 2877.6762 $\pm$ 0.0005 & $\mathcal{U}$(2853.8, 2854.0)& 2853.8872 $\pm$ 0.0005  \rule[-1ex]{0pt}{0pt}\\
   Scaled semi-major axis ($\frac{a}{R_{\star}}$) & & ... & 11.26 $\pm$ 0.17 & ... & 19.11 $\pm$ 0.29 \rule[-1ex]{0pt}{0pt}\\
   Orbital semi-major axis ($a$) & AU & ... & 0.0445 $\pm$ 0.0008 & ... & 0.076 $\pm$ 0.002 \rule{0pt}{2.0ex} \rule[-1ex]{0pt}{0pt}\\
   Orbital inclination ($i$) & deg & ... & 86.92 $\pm$ 0.14 & ... & 86.64 $\pm$ 0.17\rule{0pt}{2.0ex} \rule[-1ex]{0pt}{0pt}\\
   Orbital eccentricity ($e$) &  & 0.0 & 0.0  & 0.0 & 0.0\rule{0pt}{2.0ex} \rule[-1ex]{0pt}{0pt}\\
   Impact parameter ($b$) & & $\mathcal{U}$(0, 2) & 0.61 $\pm$ 0.02 & $\mathcal{U}$(0, 2) & 1.12 $\pm$ 0.05\rule{0pt}{1.0ex} \rule[-1ex]{0pt}{0pt}\\
   Planet/star radius ratio ($\frac{R_{\rm p}}{R_{\star}}$) & & $\mathcal{U}$(0, 0.25) & 0.0312$\pm$0.0005 & $\mathcal{U}$(0, 0.25) & 0.18 $\pm$ 0.05 \rule[-1ex]{0pt}{0pt}\\
   Argument of pericenter ($\omega$) & deg & ... & 90 & ... & 90 \rule[-1ex]{0pt}{0pt}\\
   Mean longitude ($L$) & deg & ... & 65.07 $\pm$ 0.05 & ... & 282.30 $\pm$ 0.02\rule{0pt}{2.0ex} \rule[-1ex]{0pt}{0pt}\\
   Transit duration ($T_{14}$)\tablefootmark{a} & hours & ... & 2.07 $\pm$ 0.02  & ... & 1.22 $\pm$ 0.02 \rule[-1ex]{0pt}{0pt}\\
   RV semi-amplitude ($K$) & m s$^{-1}$ & $\mathcal{U}$(0.01, 2000) & 12.8$^{+1.4}_{-1.5}$ & $\mathcal{U}$(0.01, 2000) & 12.8 $\pm$ 1.3 \rule{0pt}{2.0ex} \rule[-1ex]{0pt}{0pt}\\
   Planetary radius ($R_{\rm p}$) & $R_{\oplus}$ & ... & 2.89 $\pm$ 0.06 & ... & $>7.5~$\tablefootmark{b}  \rule[-1ex]{0pt}{0pt}\\
   Planetary mass ($M_{\rm p}$) & $M_{\oplus}$ & ... & 28.1$^{+3.3}_{-3.4}$  & ... & 36.6$^{+4.0}_{-3.9}$ \rule[-1ex]{0pt}{0pt}\\
   Planetary density ($\rho_{\rm p}$) & g cm$^{-3}$ & ... & 6.5 $\pm$ 0.9 & ... & $< 0.48$ \rule{0pt}{2.0ex}\rule[-1ex]{0pt}{0pt}\\
\hline
\hline
\multicolumn{2}{c}{Stellar parameters} & \multicolumn{4}{c}{TOI-1533} \rule{0pt}{2.3ex} \rule[-1ex]{0pt}{0pt}\\ 
\hline
Parameter & Unit & \multicolumn{2}{c}{Prior} & \multicolumn{2}{c}{Value}\rule{0pt}{2.3ex} \rule[-1ex]{0pt}{0pt}\\ 
\hline
   RV jitter & m s$^{-1}$ & \multicolumn{2}{c}{...} & \multicolumn{2}{c}{1.1$^{+1.3}_{-0.8}$} \rule{0pt}{2.3ex} \rule[-1ex]{0pt}{0pt}\\
   RV offset & m s$^{-1}$ & \multicolumn{2}{c}{...} & \multicolumn{2}{c}{$-23736.8^{+2.6}_{-3.0}$} \rule{0pt}{2.3ex} \rule[-1ex]{0pt}{0pt}\\
   Density ($\rho_{\star}$) & $\rho_{\odot}$ & \multicolumn{2}{c}{$\mathcal{N}$(1.44, 0.07)} & \multicolumn{2}{c}{$1.44 \pm 0.07$}   \rule{0pt}{2.3ex} \rule[-1ex]{0pt}{0pt}\\
   Rotation period ($P_{\rm rot}$) & days & \multicolumn{2}{c}{$\mathcal{U}$(10, 40)} & \multicolumn{2}{c}{20.02$^{+0.37}_{-0.33}$} \rule{0pt}{2.3ex} \rule[-1ex]{0pt}{0pt}\\
   Decay Timescale of activity ($P_{\rm dec}$) & days & \multicolumn{2}{c}{$\mathcal{U}$(20, 1000)} & \multicolumn{2}{c}{28.1$^{+8.0}_{-5.3}$} \rule{0pt}{2.3ex} \rule[-1ex]{0pt}{0pt}\\
   Coherence scale ($\omega_{\rm cs}$) &  & \multicolumn{2}{c}{$\mathcal{N}$(0.350, 0.035)\tablefootmark{c}} & \multicolumn{2}{c}{0.37$\pm$0.03} \rule{0pt}{2.3ex}   \rule[-1ex]{0pt}{0pt}\\
   \hline
\end{tabular}
 }
 \tablefoot{\tablefoottext{a}{From \cite{2010exop.book...55W}.} \tablefoottext{b}{0.13th percentile lower limit.} \tablefoottext{c}{e.g. \cite{2022A&A...664A.163N}.}}
\end{table*}

\subsection{Simultaneous photometry as part of the multidimensional GP framework}

Simultaneous photometric (\textit{TESS}) and spectroscopic (HARPS-N) observations have been fundamental to disentangling Keplerian signals from stellar activity. In particular, sectors 84 and 85 of \textit{TESS} covered most of the HARPS-N observation time span, enabling more than one stellar rotation period to be modelled simultaneously alongside the spectroscopic activity indicators. We included the simultaneous photometry as one of the observables in the multidimensional GP formalism introduced by \cite{2015MNRAS.452.2269R} but did not use its first derivative. However, after careful inspection (further detailed in Appendix \ref{app:gp}) of the TESS photometry, we decided not to include sector 85 in the multidimensional GP. We hence performed the analysis with a four-dimensional GP model as follows:

\begin{equation}
\begin{split}
\label{eq:model}
    \Delta RV & = V_{\rm c}\,G(t) + V_{\rm r}\,\dot{G}(t)~,\\
    {\rm FWHM} & = L_{\rm c}\,G(t)~,\\
    \log R'_{\rm HK} & = L2_{\rm c}\,G(t)~,\\
    Flux & = L3_{\rm c}\,G(t)~,
\end{split}
\end{equation} where $G(t)$ and $\dot{G}(t)$ are the GP and its time derivative, while $V_{\rm c}, V_{\rm r}, L_{\rm c}, L2_{\rm c}, L3_{\rm c}$ coefficients characterise the amplitude of $G(t)$ and $\dot{G}(t)$ for each time series \citep[see e.g.][]{2022MNRAS.509..866B}. 

\begin{figure}
   \centering
   \sidecaption
   \includegraphics[width=0.88\hsize]%
   {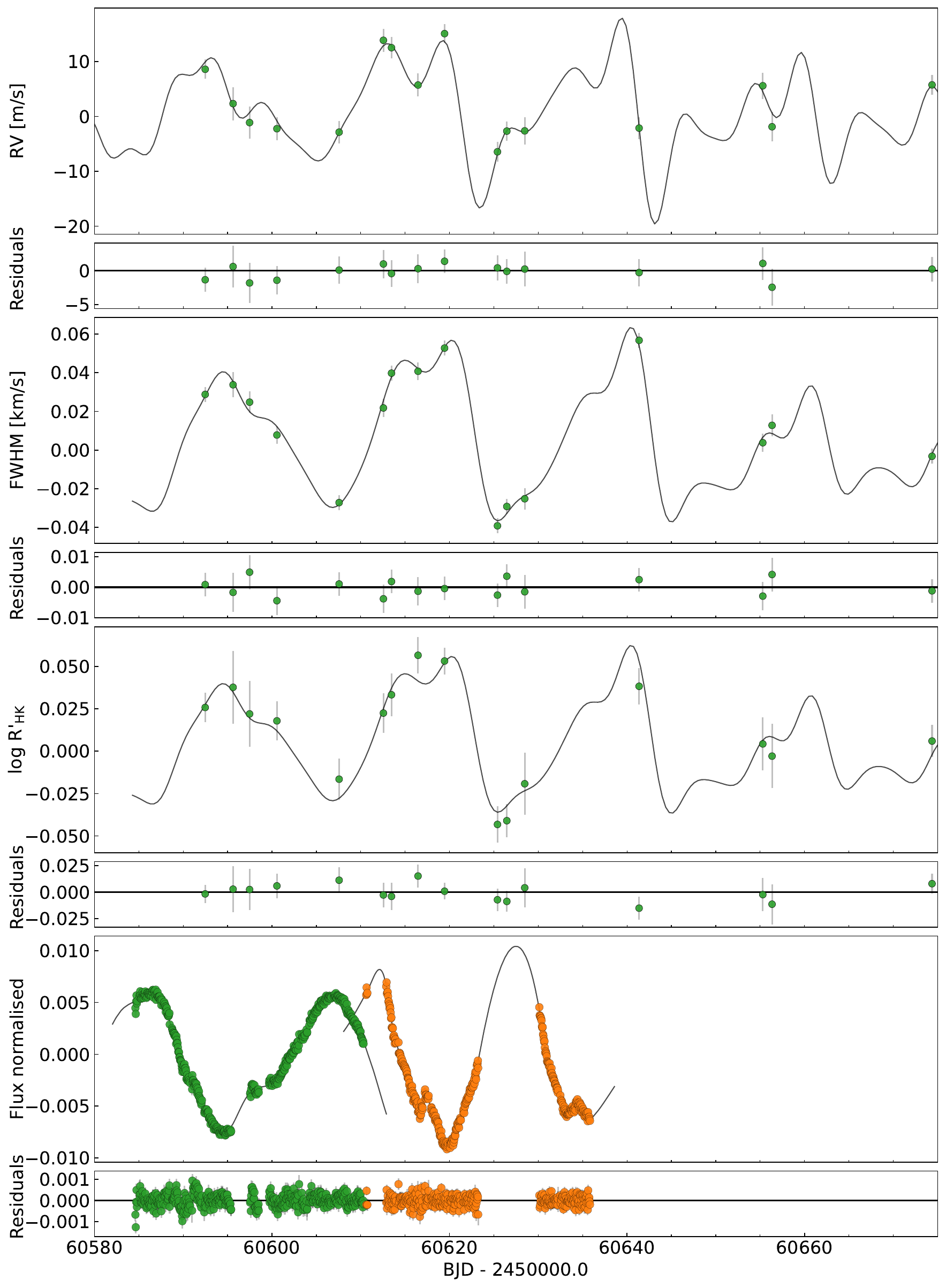}
   \caption{Multi-dimensional GP model of stellar activity in the RV, FWHM, $\log R'_{\rm HK}$, and photometric time series. Each panel shows the time series modelling and the residuals of the fit. \textit{Top panel:} RV time series without Keplerian signals. \textit{Bottom:} Photometric time series of simultaneous \textit{TESS} sectors 84 and 85. Sector 85 is shown in a different colour because it was not included in the multidimensional GP framework; instead, it was modelled with a unidimensional GP (see Appendix \ref{app:gp} for further clarification).}
   \label{fig:multigp}
\end{figure}

Table \ref{tab:gps} shows a comparison of the case in which we considered the simultaneous photometry as part of the multidimensional modelling with the case in which we did not. It is worth noting that while the amplitudes of both planetary signals remain similar, the RV jitter is reduced by more than 50\% when we consider simultaneous photometry as part of the model. When we included the photometry, the GP amplitudes of the RV datasets remained statistically consistent, but their uncertainty decreased. The uncertainty in the stellar hyperparameters and coefficients also decreased. This inclusion provided a better constraint on all the GP hyperparameters, which in turn reduced the RV jitter. We therefore adopted the latter model as the reference (see Fig. \ref{fig:multigp}).

Lastly, we want to emphasise that we repeated the analysis, allowing the eccentricity to vary. However, our analysis showed a clear preference for case 1 (circular) over case 2 (eccentric), with a significant $\Delta {\rm BIC_{21}}$ value of 64 \citep{doi:10.1080/01621459.1995.10476572}. Even when only the eccentricity of the inner planet was allowed to vary (case 3), case 1 was still favoured, with a significant $\Delta {\rm BIC_{31}}$ of 20. The same is true when allowing the eccentricity of the outer planet to vary (case 4), with a significant $\Delta {\rm BIC_{41}}$ of 18. 

\begin{table}[t]
\tiny
\caption{Comparison of the hyperparameters of the two multidimensional GPs used to model the stellar activity.}  
\label{tab:gps}
\centering   
\begin{tabular}{l | c | c}     
\hline\hline         
 Parameter & w/o photometry & w photometry \rule{0pt}{2.5ex} \rule[-1ex]{0pt}{0pt} \\
\hline         
    $\sigma^{\rm RV}_{\rm jitter}$ (m s$^{-1}$) & 3.3$^{+2.3}_{-2.1}$ & 1.1$^{+1.3}_{-0.8}$ \rule{0pt}{2.5ex} \rule[-1ex]{0pt}{0pt} \\
    $K_{\rm b}$ (m s$^{-1}$)  & 12.3$^{+2.2}_{-2.1}$ & 12.8$^{+1.4}_{-1.5}$ \rule{0pt}{2.5ex} \rule[-1ex]{0pt}{0pt} \\
    $K_{\rm c}$ (m s$^{-1}$)  & 12.8$^{+2.2}_{-2.1}$ & 12.8$^{+1.3}_{-1.3}$ \rule{0pt}{2.5ex} \rule[-1ex]{0pt}{0pt} \\
    $P_{\rm rot}$ (days)  & 20.66$^{+0.62}_{-0.41}$ & 20.02$^{+0.37}_{-0.33}$ \rule{0pt}{2.5ex} \rule[-1ex]{0pt}{0pt} \\
    $P_{\rm dec}$ (days)  & 52.3$^{+20}_{-8.3}$ & 28.1$^{+8.0}_{-5.3}$ \rule{0pt}{2.5ex} \rule[-1ex]{0pt}{0pt} \\
    $V_{\rm c}$ (m s$^{-1}$)  & 7.6$^{+3.6}_{-2.9}$ & 5.8$^{+2.1}_{-1.5}$ \rule{0pt}{2.5ex} \rule[-1ex]{0pt}{0pt} \\
    $V_{\rm r}$ (m s$^{-1}$)  & 19.1$^{+9.7}_{-8.5}$ & 18.3$^{+5.8}_{-4.3}$ \rule{0pt}{2.5ex} \rule[-1ex]{0pt}{0pt} \\
    $L_{\rm c}$ (km s$^{-1}$)  & 0.04$^{+0.02}_{-0.01}$ & 0.03$\pm$0.01 \rule{0pt}{2.5ex} \rule[-1ex]{0pt}{0pt} \\
    $L2_{\rm c}$   & 0.04$^{+0.02}_{-0.01}$ & 0.03$\pm$0.01 \rule{0pt}{2.5ex} \rule[-1ex]{0pt}{0pt} \\
\hline                                   
\end{tabular}
\tablefoot{We omitted the coherence scale ($\omega_{\rm cs}$) as its value is mainly driven by the prior.}
\end{table}

\subsection{Joint modelling with NEID data}
We included the NEID RVs in the simultaneous photometric and spectroscopic analysis using \texttt{PyORBIT}, but there was no improvement compared to using only the HARPS-N and \textit{TESS} data. While the NEID RVs do agree with the stellar activity model, the uncertainty in the activity coefficients is greater, most likely because the seven NEID data points are too sparse and widely separated in time from the HARPS-N RVs. This made it difficult to properly model the stellar activity contribution in the NEID data.

Nevertheless, the NEID data are an important addition to the analysis presented in the previous section. In particular, the peak-to-peak RV variation is comparable in both datasets (see Figs. \ref{fig:rv_full} and \ref{fig:rv_full_neid}), which confirms the significance of the activity contribution and the two Keplerian signal amplitudes. Moreover, this result highlights that the GP we used is robust and did not overfit the data, which is significant given the difficulty in modelling the two widely separated datasets.

\subsection{Search for transit timing variations}
We explored the possible presence of dynamical interactions in the system by performing a search for transit timing variations \citep[TTVs, e.g.][]{2005MNRAS.359..567A, 2005Sci...307.1288H} of both planets using a \texttt{PyORBIT} model based on the \texttt{BATMAN} code. We fitted each $T_0$, fixing the orbital periods described in Sect. \ref{sec:analysis}.

We prepared the observed (O) minus calculated (C) diagrams by removing the linear ephemeris for each $T_0$. The O--C diagrams are presented in Figs. \ref{fig:ttv_b} and \ref{fig:ttv_c}. Interestingly, for the inner planet TOI-1533 b, the possible TTV amplitude ($A_{\rm TTV}$), computed as the semi-amplitude of the O--C, is $24 \pm 7$ minutes \citep[with the uncertainty estimated as in ][]{2024A&A...691A..67M}. The reduced $\chi^2$ value of the linear ephemeris is about seven, which suggests ongoing dynamical interaction between planets b and c -- likewise recently suggested by \cite{2026arXiv260617218N}. 

Motivated by this finding, we performed a dynamical analysis using the \textit{N}-body dynamical integrator \texttt{TRADES}\footnote{\url{https://github.com/lucaborsato/trades}}\citep{2014A&A...571A..38B, 2019MNRAS.484.3233B, 2021MNRAS.506.3810B}. While our analysis (see Appendix \ref{app:dyn}) successfully retrieved planetary parameters consistent with those reported in Table \ref{table:model-lcrv}, our dynamical model is not capable of reproducing the observed transit times with a total reduced $\chi^2$ of seven for the best-fit configuration. The residuals could be the results of low-accuracy and low-precision observations, additional body (or bodies) in the system, or in-transit stellar activity in the form of stellar spots that affect the \textit{TESS} data. Only exquisite, high-precision photometry, such as that provided by the CHaracterising ExOPlanet Satellite \citep[CHEOPS,][]{2021ExA....51..109B}, could determine if the observed variation is of planetary or stellar origin.

\subsection{Planet equilibrium temperature and TSM}
\label{sec:tsm}
We estimated the equilibrium temperature of both planets, assuming zero albedo and full day-night heat redistribution, with the simple equation $T_{\rm eq} = T_{\rm eff} \sqrt{\frac{R_\star}{a}}\left(\frac{1}{4}\right)^{1/4}$, with $T_{\rm eff}$ as the stellar effective temperature, $R_\star$ as the stellar radius, and $a$ as the orbital semi-major axis. We found $T_{\rm eq} = 1084 \pm 20$ K for TOI-1533 b and $T_{\rm eq} = 829 \pm 17$ K for planet c.

Therefore, we derived the transmission spectroscopy metric \citep[TSM;][]{2018PASP..130k4401K} for planet b to be $20 \pm 3$ by inserting the estimated equilibrium temperature into the TSM equation. In contrast, due to the grazing nature of TOI-1533 c (see Table \ref{table:model-lcrv}), we were unable to obtain a precise TSM estimation for the outer companion. Nevertheless, based on the minimum retrieved radius, we could set a minimum TSM value of about 200. However, to compensate for the fact that the signal-to-noise ratio of an atmospheric observation scales as the square root of the transit duration, we adjusted the TSM to account for the grazing nature of TOI-1533~c. We estimated the scaling factor as $\sqrt{T_{14,~\rm b>1} / T_{14,~\rm b=0}} = \sqrt{1.22/3.8} = 0.566$, where $T_{14,~\rm b=0} \approx \frac{P}{\pi}\frac{R_\star}{a}$ from \cite{2026A&A...710A.300N}, implying a corrected TSM$_{\rm corr}$ of approximately $ 110$. Therefore, even with this correction, the minimum TSM value implies that TOI-1533~c is a high-quality atmospheric characterisation target.

\subsection{Ephemeris improvements and future observations}
Propagating the new ephemeris to 1 January 2030 reduces the level of uncertainty for TOI-1533~b to 4.2 minutes and for TOI-1533~c to 3.4 minutes. At present, no further \textit{TESS} observations are planned for this target.

\section{Discussion}
\label{sec:discussion}

\subsection{A new compact multi-planet system with a hot gas giant}

This paper confirms TOI-1533 as one of only a handful of multi-planet systems with a short-period giant planet and an inner small-size planet companion. Such multi-planet systems are an emerging category of planetary system architecture \citep[see e.g.][]{2025AJ....169..149H} that can help provide better understanding of formation and migration mechanisms. They also play a key role in linking the exoplanet population to the architecture and dynamics of the Solar System.

\begin{figure}
   \centering
   \includegraphics[width=0.8\hsize]%
   {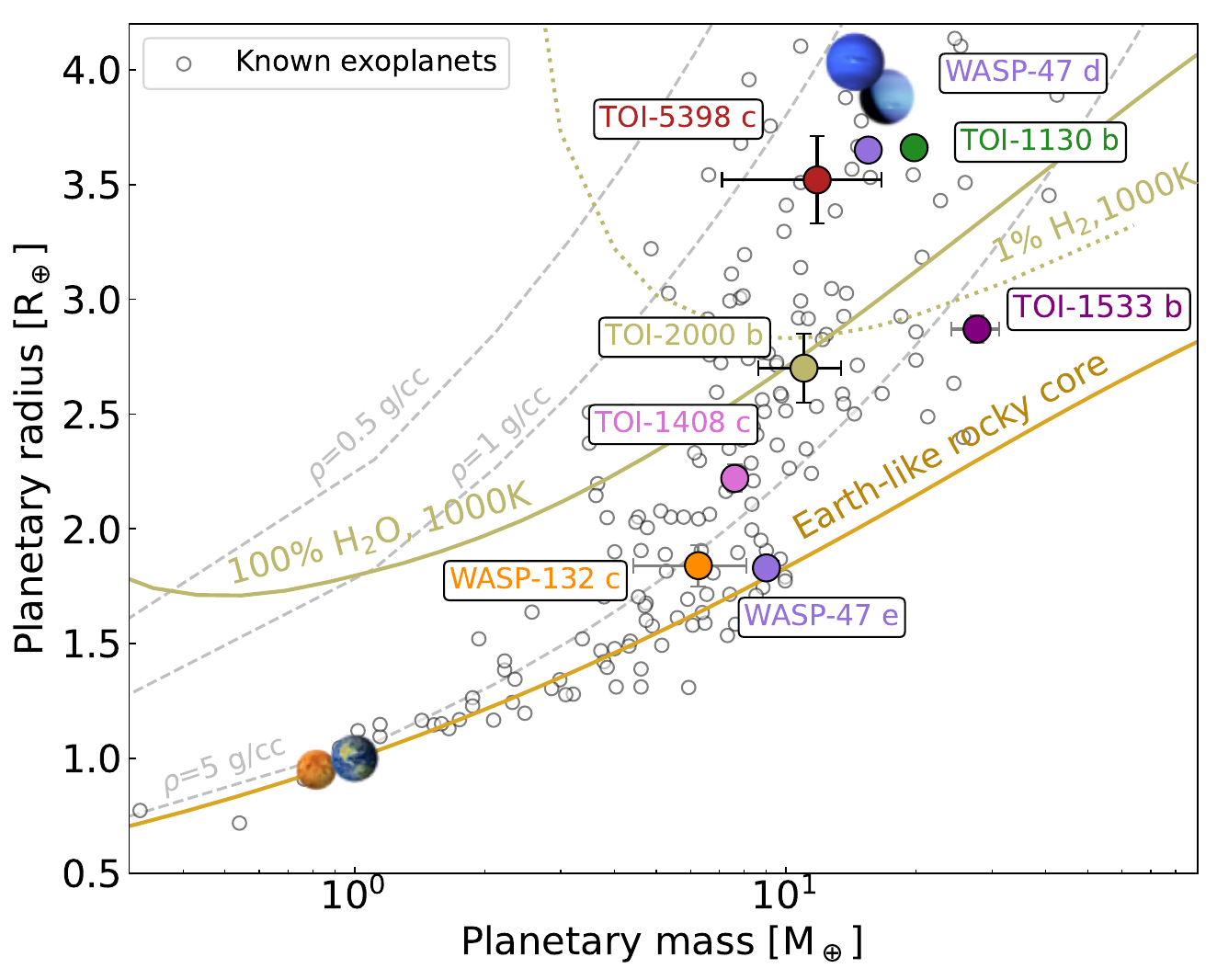}
   \caption{Mass–radius diagram of all confirmed small planets ($R_{\rm p} < 4 R_\oplus$) with mass and radius precisions better than 20\% \citep{2025PSJ.....6..186C}. The masses of the plotted small inner companions of hot giants have been taken from references listed in Tab. \ref{tab:companions}. We included theoretical mass--radius curves for various planetary pure compositions from \cite{2019PNAS..116.9723Z}. Grey dashed curves show densities $\rho$ = 0.5, 1, and 5 g~cm$^{-3}$.}
   \label{fig:mass-radius}
\end{figure}

TOI-1533 b is a transiting sub-Neptune with a radius of $R_{\rm p} = 2.89 \pm 0.06~R_\oplus$ and a mass of $M_{\rm p} = 28.1^{+3.3}_{-3.4}~M_\oplus$ that orbits a K-dwarf star in about 3.6 days. Its outer companion, TOI-1533 c, is a grazing giant planet with a minimum radius of 7.5 $R_\oplus$ (0.13th percentile lower limit; see also Fig. \ref{fig:post_distr}) and a mass of $M_{\rm p} = 36.6^{+4.0}_{-3.9}~M_\oplus$, about one third that of Saturn. From the minimum radius, we can derive a maximum planetary density of $\rho_{\rm p} < 0.48$ g cm$^{-3}$, which is compatible with a hot gas giant planet primarily composed of H/He by volume and mass \citep[e.g.][]{2014ApJ...792....1L, 2019PNAS..116.9723Z}. Compared to small-size planets orbiting interior to short-period giant planets (see Fig. \ref{fig:mass-radius}), TOI-1533~b is instead the most massive small-size planet, with a mass almost double that of Neptune and about 75\% of its outer giant companion. Moreover, TOI-1533~b joins a rare subpopulation of Neptunian planets with high densities \citep[see][]{2023MNRAS.526..548O, 2020Natur.583...39A, 2023Natur.622..255N}. Notably, it is the only one of these high-density planets known to be in a multi-planet system.

\subsection{Architecture of systems hosting hot, intermediate-mass planets}

\begin{figure}
   \centering
   \includegraphics[width=0.8\hsize]%
   {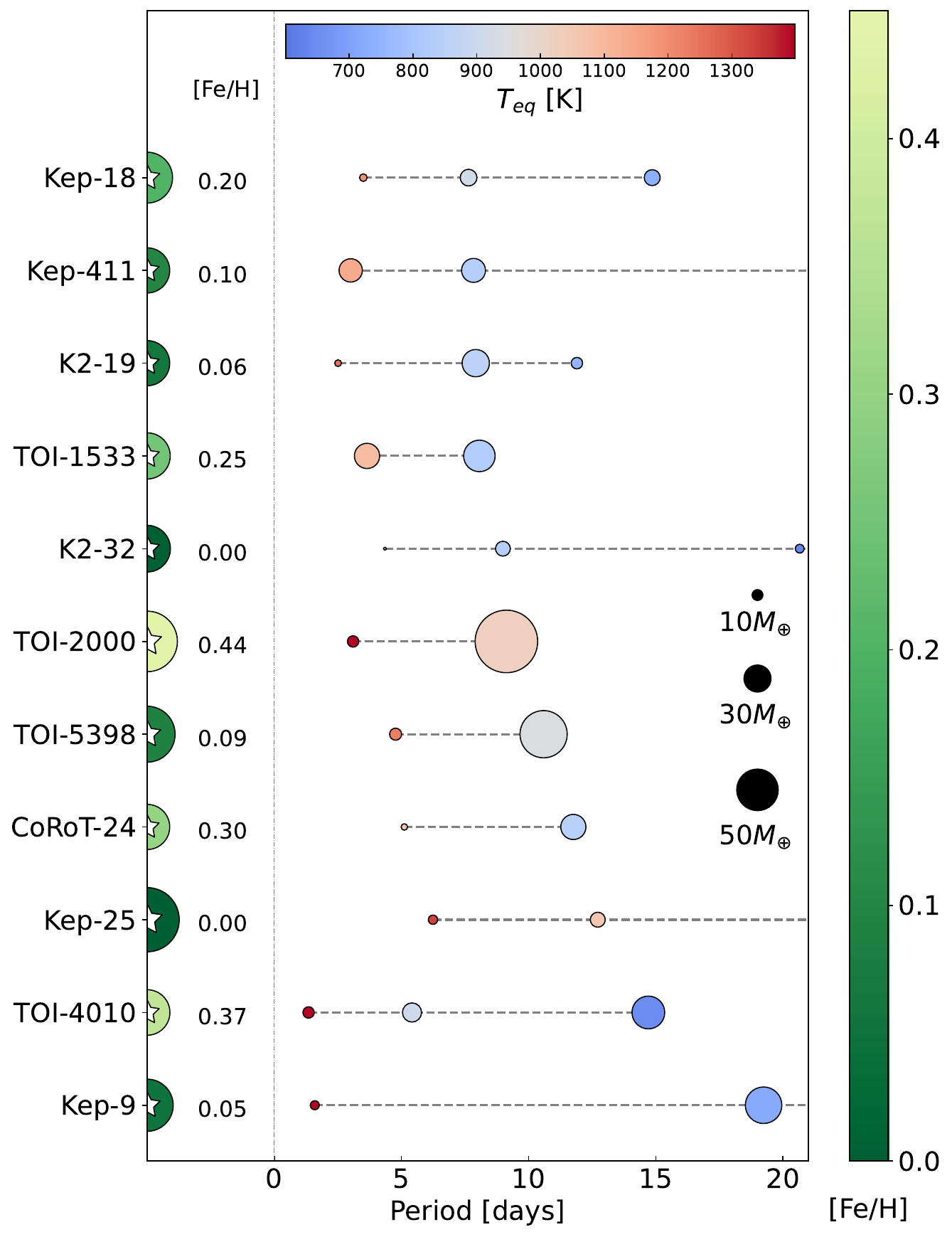}
   \caption{Architecture of multi-planet systems hosting an intermediate-mass planet ($15 M_\oplus < M_{\rm p} < 120 M_\oplus$, $R_{\rm p} > 4 R_\oplus$) and an inner companion, both in close-in orbits ($P < 20$~d). Each row represents a planetary system (vertical axis) and the planetary orbital periods (horizontal axis). The systems are sorted in ascending order of the period of the intermediate-mass planet. The equilibrium temperature of the planets is colour-coded (top colour bar), while the dot size tracks planetary masses. The host stars are indicated by semi-circular dots, colour-coded by stellar metallicity (right colour bar) and with sizes that encode their radii. Dashed lines indicate the outermost planets not shown.}
   \label{fig:saturn_neptune}
\end{figure}

The mass ratio of the components of this system, together with the low mass of the large-size outer companion, is what makes it quite unusual compared to classical HJs with inner low-mass companions. Motivated by these findings, we selected all the confirmed planets \citep{2025PSJ.....6..186C} similar to TOI-1533 c and that have inner companions of any kind. Our selection criteria -- for the outer planet -- were the following:
\begin{itemize}
    \item Orbital period shorter than 20 days.
    \item Planetary radius $R_{\rm p} > 4~ R_\oplus$.
    \item Planetary mass ($M_{\rm p}$) between 15 $M_\oplus$ and 120 $M_\oplus$, to include Neptunes and Saturns.
\end{itemize} We found a total of eleven systems (Fig. \ref{fig:saturn_neptune}, Tab. \ref{tab:companions}). One interesting trend that emerged is the preponderance of metal-rich stars. The only two stars with solar metallicity (Kepler-25, K2-32) host the two least massive planets (about 15 $M_\oplus$ each). This common characteristic becomes even more intriguing if we include the other six compact systems with short-period giants (see Tab. \ref{tab:companions}), as they also orbit metal-rich stars. This result follows the well-known trend that planets are found more frequently around metal-rich stars \citep{2001A&A...373.1019S, 2004A&A...415.1153S}. It is particularly evident for giant planets, as the likelihood of their formation increases around such stars, as predicted by the core-accretion model \citep{2008ASPC..398..235M}. However, it is rather peculiar that none of the selected systems is hosted by a metal-poor star. By contrast, about 20\% of single-planet systems with these characteristics orbit metal-poor stars.

\begin{figure}
   \centering
   \includegraphics[width=0.73\hsize]%
   {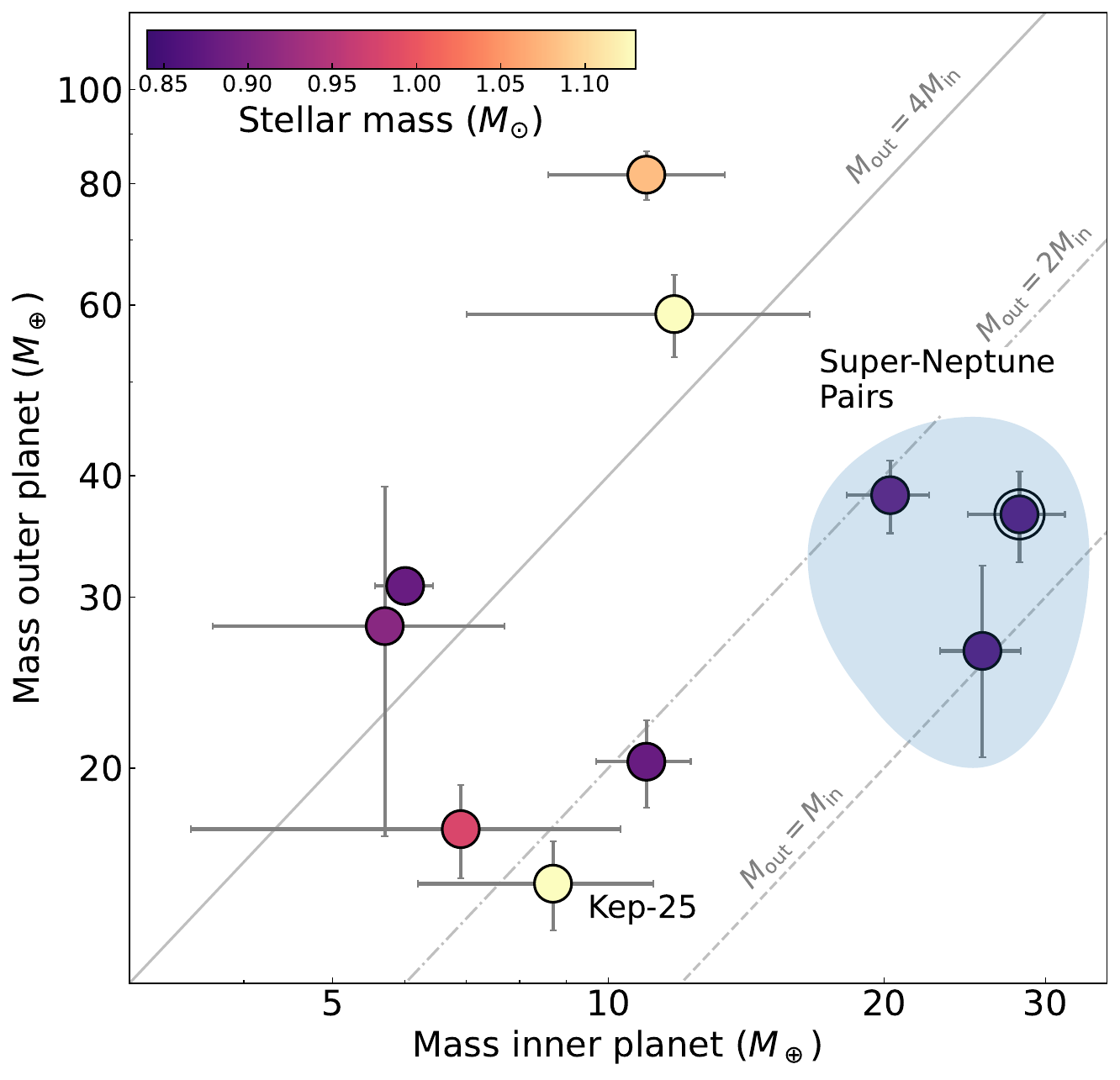}
   \caption{Inner planet mass vs outer planet mass for all the systems in Fig. \ref{fig:ratios}. The stellar mass is colour-coded, while the blue area highlights planet pairs with similar masses. TOI-1533 is indicated by two black circles. }
   \label{fig:masses}
\end{figure}

Figure \ref{fig:saturn_neptune} seems to also suggest that stars with larger radii and masses (F-G dwarfs) tend to have hot Saturn planets in terms of masses, while smaller stars (K dwarfs) seem to host less massive planets, such as hot Neptunes and super-Neptunes (see Fig. \ref{fig:masses}). Also in this case, Kepler-25 appears unusual, as it is an F-type star hosting the lowest-mass outer planet in the sample. However, the system also hosts a 120-day Saturn-mass planet. In Fig. \ref{fig:masses}, we highlight pairs of planets with similar masses, including TOI-1533. Interestingly, the sizes, masses, and periods of the planets orbiting Kepler-411 are comparable to those of TOI-1533, with the inner planet being primarily rocky and the outer one having gases dominating its volume. However, Kepler-411 also has two known planets at longer orbital periods. We explore this further in Sect. \ref{sec:outer}.

\subsection{Migration history implications}

\begin{figure}
   \centering
   \includegraphics[width=0.8\hsize]%
   {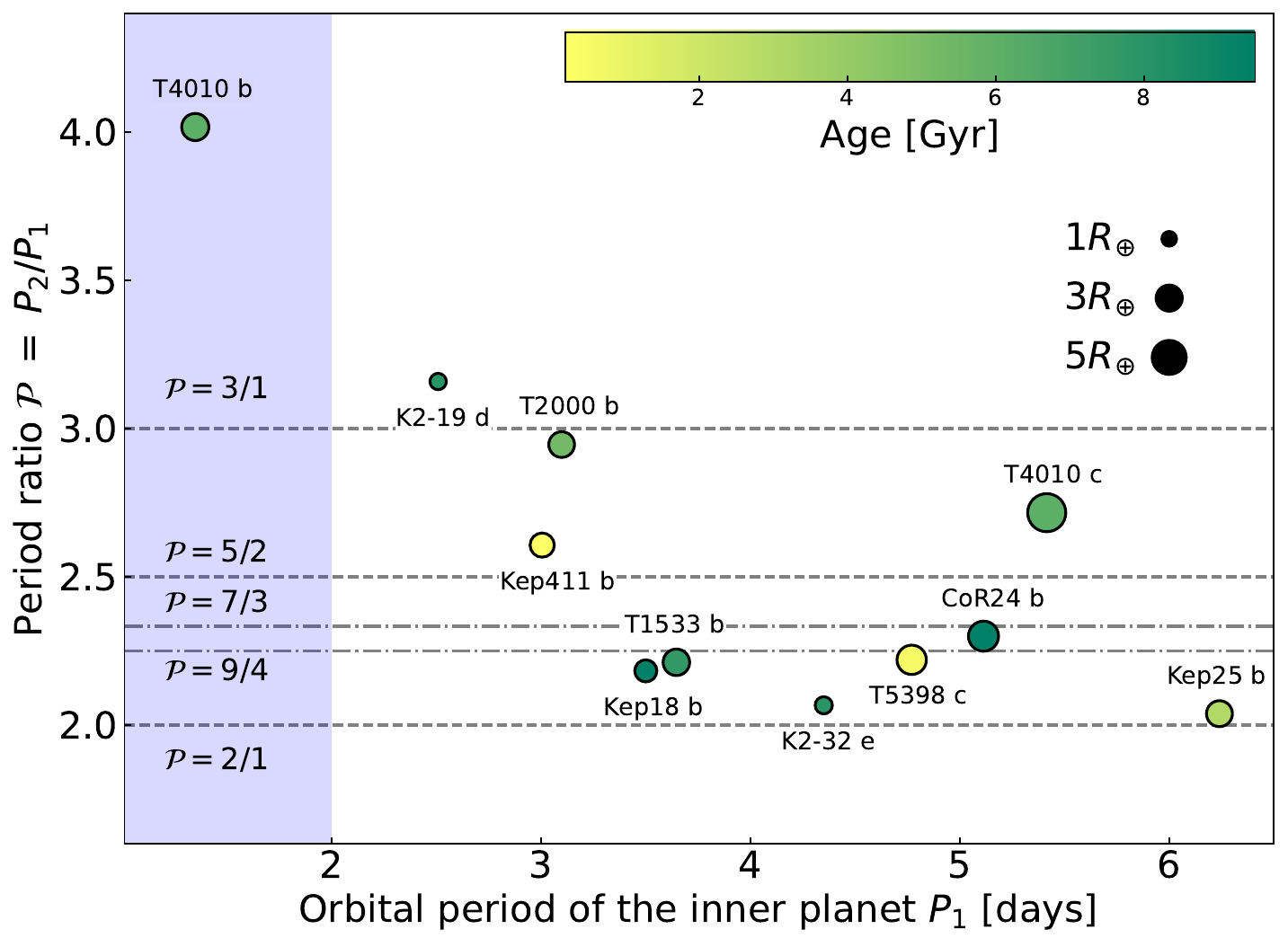}
   \caption{Orbital period $P_1$ as a function of the period ratio $\mathcal{P} = P_2/P_1$ for all planets interior to close-in intermediate-mass planets ($15 M_\oplus < M_{\rm p} < 120 M_\oplus$, $R_{\rm p} > 4 R_\oplus$). The age of the system is colour-coded, and the dot size tracks the radius of the planet it represents. Horizontal lines represent some low-order mean-motion resonances, and the blue area highlights planets with orbits shorter than two days. }
   \label{fig:ratios}
\end{figure}

In light of their possible shared disc-driven migration history, we inspected whether the selected multi-planet systems exhibit similar characteristics. As shown in simulations by \cite{2006Sci...313.1413R}, at early stages of planetary formation, systems with a planet orbiting interior to a close-in giant often cluster near mean-motion resonances (i.e. where the period ratio $\mathcal{P} \equiv P_2/P_1$ can be expressed as a rational number $i/j$), where gravitational interactions can lead to detectable TTVs. However, the resonant configuration can be disrupted by disc dissipation \citep[e.g.][]{2023A&A...679A..55T} or diverging planet evolution at late times. The latter is likely to result from either mass loss due to photoevaporation or from different tidal interaction strengths with the star, causing the inner companion to detach and increase $\mathcal{P}$. In Fig. \ref{fig:ratios}, we show the orbital period of the inner planets ($P_1$) as a function of the planet pair period ratio ($\mathcal{P}$). There is a clear upward trend in the values of $\mathcal{P}$ towards shorter orbital periods, with the largest period ratios observed for planets within two days (TOI-4010~b and Kepler-9~d, the latter being even outside the y-axis range), which is in line with a very recent finding by \cite{2025AJ....169..191G}. It is worth noting that planets with a larger $\mathcal{P}$ but similar $P_1$ tend to have larger $R_{\rm p}$ values.

\begin{figure}
   \centering
   \includegraphics[width=0.7\hsize]%
   {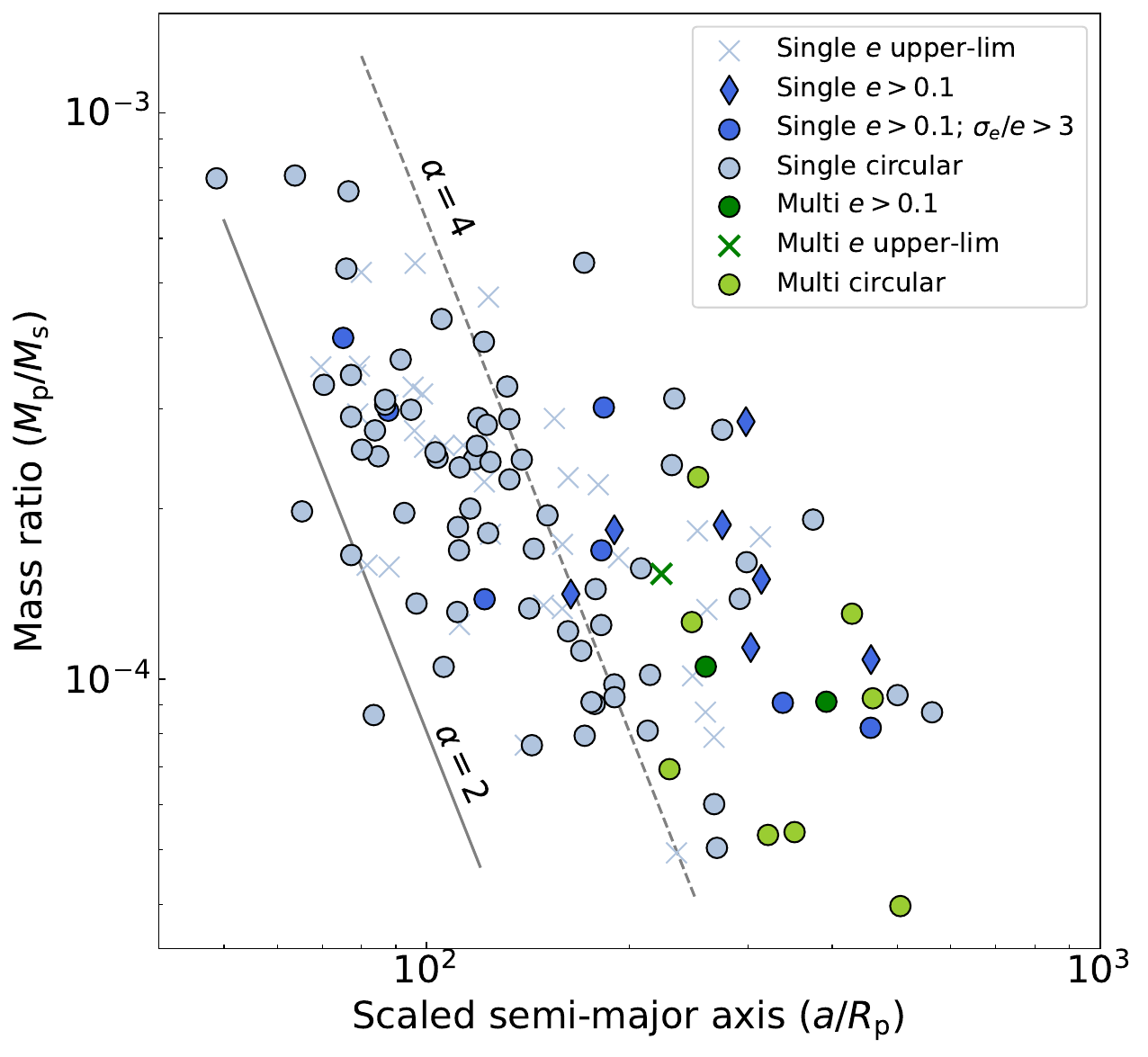}
   \caption{Tidal diagram of intermediate-mass planets in single- and close-in multi-planet systems. Single-planet systems are coloured light blue (circular) and blue (eccentric), with different shapes representing their eccentricity precision. Multi-planet systems are coloured green (eccentric) and light green (circular). The solid and dashed lines respectively indicate $a = 2 a_{\rm R}$ and $a = 4 a_{\rm R}$, with $a_{\rm R}$ as the Roche limit. }
   \label{fig:a_mass}
\end{figure}

Planets undergoing disc-driven migration may share two common characteristics that influence their natural photoevaporation resistance. First, they arrive in close-in orbits in a short time -- $\lesssim$ 10 Myr -- exposing them to full, constant extreme ultraviolet (XUV) irradiation during the stellar active phase. This contrasts with planets undergoing late high-eccentricity migration, which can arrive close-in much later and may have escaped this active phase \citep{2018Natur.553..477B, 2021A&A...647A..40A}. Another possible driver of atmospheric erosion is the lower atmospheric metallicity and C/O ratios foreseen for planets undergoing disc-migration \citep{2014ApJ...791L...9M,2024MNRAS.535..171P, 2026MNRAS.546ag143C}. This may in turn make these planets less resistant to photoevaporation, resulting in greater mass loss \citep[e.g.][]{2022MNRAS.511.1043W, 2024A&A...691A..67M}. Motivated by these two characteristics, we compared Neptune-mass planets with inner companions to those in single-planet systems (or whose companions are outer long-period ones). To this end, we generated a tidal diagram of scaled semi-major axis ($a/R_{\rm p}$) versus mass ratio ($M_{\rm p}/M_{\rm s}$) for all Neptune-mass planets in single and multi-planet systems (Fig. \ref{fig:a_mass}) while also distinguishing between circular and eccentric orbits.  

Planets in close-in multi-planet systems tend to have larger scaled semi-major axes -- which is possibly due to the inner companion hindering the inward migration of the outer planet -- and potentially lower masses. This possible trend may indicate greater atmospheric erosion for planets undergoing disc-driven migration. It is also worth noting that almost all planets with $\alpha$ $< 4$ are on circular orbits, as would be expected given that the timescale for tidal circularisation increases with increasing $\alpha$ values \citep{1966Icar....5..375G, 2017A&A...602A.107B}.

\subsection{Prospects for atmospheric follow-up}

The unusual architecture of this system provides a valuable laboratory for investigating the interplay between planet formation and atmospheric composition. The atmospheres of the planets in this system are expected to preserve chemical fingerprints of their accretion histories, encoding information about where and how material was accreted within the protoplanetary disk \citep{2011ApJ...743L..16O}. Since disk chemistry varies with radial distance, particularly across major ice lines, different solid-to-gas accretion ratios imprint distinct elemental and molecular signatures on planetary atmospheres. In particular, the relative abundances of water- and carbon-bearing species as well as the ratio of volatiles to refractories may provide key constraints on formation location and migration history \citep{2021ApJ...914...12L,2022ApJ...937...36P, 2023ApJ...943..112C}. 

In the TOI-1533 system, the orbital architecture will strongly influence the atmospheric properties of the smaller inner planet. The presence of an outer gas giant companion is expected to regulate or inhibit the inward drift of icy pebbles \citep{2024ApJ...977...21E}, thereby altering the volatile budget of the inner disk where TOI-1533 b formed. Such a mechanism would directly affect the planet’s water enrichment and, consequently, its observable abundance ratios, such as the carbon-to-oxygen ratio (C/O), potentially favouring a higher C/O ratio. 

The most observationally accessible planet in the system is TOI-1533 c (TSM $\gtrsim$ 200 and TSM$_{\rm corr} \approx 110$, compared to TSM $\approx$ 20 for TOI-1533 b). Despite its grazing nature, this high TSM places the outer giant among the most favourable targets for atmospheric characterisation at its equilibrium temperature (Fig. \ref{fig:TSM_TOI_1533c}, where we did not use the corrected TSM value to ensure consistency with all literature values). Notably, its equilibrium temperature lies near the $\mathrm{CH_4}$/CO chemical transition regime, making it particularly compelling for probing carbon chemistry \citep{2020AJ....160..288F}. Infrared observations with the James Webb Space Telescope or ground-based facilities such as IGRINS-2 would be particularly powerful for constraining the relative abundances of methane and carbon monoxide, key tracers of the atmospheric C/O ratio and disequilibrium processes (Fig. \ref{fig:simulated_spectra_TSM_TOI_1533c}). At these temperatures, vertical mixing is expected to drive significant departures from equilibrium chemistry \citep{2011ApJ...738...72V}. This is the most evident in methane, a species highly sensitive to quenching, as its abundance depends strongly on pressure and temperature. Simultaneous constraints on $\mathrm{CH_4}$ and CO would therefore provide valuable insight into the planet’s internal energy budget and vertical mixing efficiency. Further prospects are detailed in Appendix \ref{app:further}.

\subsection{Search for outer companions}
\label{sec:outer}

\begin{figure}
   \centering
   \includegraphics[width=0.92\hsize]%
   {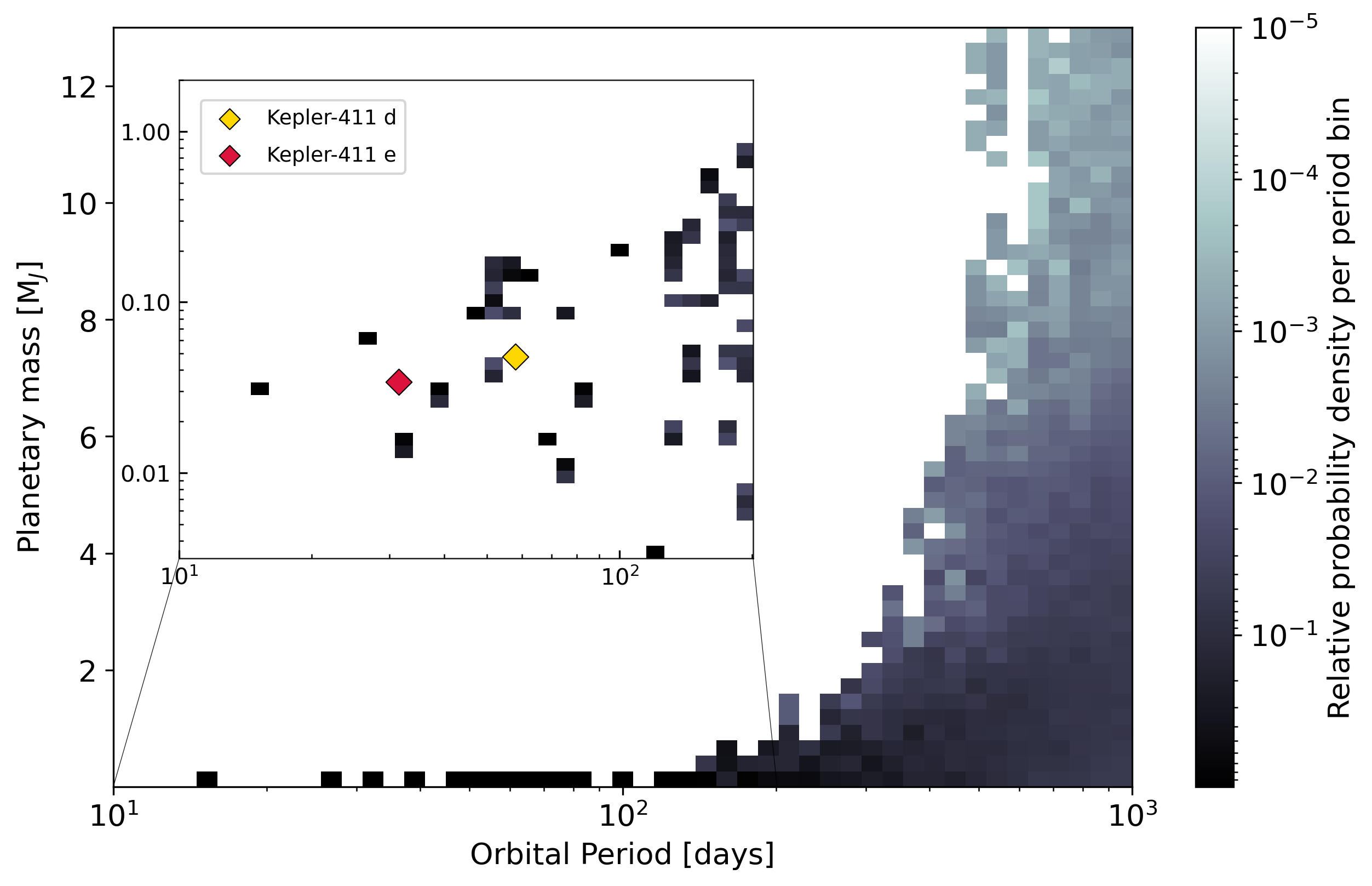}
   \caption{Posterior-based mass–period detection map for a hypothetical additional planet in the TOI-1533 system. Darker regions indicate mass–period combinations that are compatible with the residuals of the two-planet model, while brighter regions are strongly disfavoured. The short-period regime is shown in detail in the logarithmic-scale inset, where the outer planets of Kepler-411 are superimposed for comparison.}
   \label{fig:det}
\end{figure}

Besides confirming TOI-1533 b and c, we also investigated the compatibility of the data with the presence of an additional planet in the system using a posterior-based mass–period detection map (Fig. \ref{fig:det}). To generate this map, we repeated the analysis described in Sect. \ref{sec:analysis} but included an additional Keplerian component with respect to our reference model to represent a hypothetical undetected signal in the data.
Wide, uninformative priors were adopted for its orbital period ($P_{\rm orb}$) and the radial-velocity semi-amplitude ($K$). 
The posterior samples obtained for this additional component were then used to explore the region of the parameter space compatible with the residuals of the adopted two-planet model. Unlike classical injection-recovery experiments, this approach directly exploits the posterior distributions, naturally preserving only physically allowed solutions. We then computed a two-dimensional histogram of the posterior samples in the mass–period plane using 50 bins for each parameter over the range [10–1000 d, $1M_\oplus$–$13M_J$] and normalised the distribution within each period bin to obtain the conditional posterior density $P(M_{\rm p}|P_{\rm orb})$. 

Figure \ref{fig:det} shows that short-period signals ($P \lesssim 200$ d) are strongly disfavoured across the entire mass range, with relative probability densities typically below $10^{-4}$ per period bin. The only exceptions consist of a small number of sparse solutions in the sub-giant planet regime ($M_{\rm p} \lesssim 1M_J$), which likely correspond to degenerate eccentric configurations producing quasi-flat signals over the observed RV time series. Overall, the result reported in the figure supports our two-planet model, which includes the GP treatment of stellar activity, and demonstrates that it provides a robust description of the RV variability on these orbital timescales. However, as shown by the inset comparing the outer planets of Kepler-411, the presence of additional planets in this regime around TOI-1533 cannot be definitively excluded. 

At longer orbital periods, the ability of the current dataset to exclude additional companions progressively decreases. Above 200 days, only massive companions with masses of a few to several Jupiter masses can be confidently ruled out, while at even longer periods, such companions remain compatible with the observations. This behaviour reflects the limited temporal baseline of the available RV data. Additional long-term RV monitoring and \textit{Gaia} DR4 astrometry will be essential to further constraining the outer system architecture and searching for potential long-period companions.

\section{Conclusions}
\label{sec:conclusions}
In this work, we have presented the discovery of TOI-1533, a multi-planet system composed of a super-Neptune-mass pair with disparate radii orbiting an active K-dwarf star. Through the joint modelling of stellar and planetary signals from HARPS-N RVs and TESS transits, we were able to measure the planetary masses and isolate the stellar modulation. We highlight a few notable findings that emerged from this study:
\begin{itemize}
    \item TOI-1533 is confirmed as the ninth multi-planet system comprising a close-in gas giant planet ($R_{\rm p} > 7.5 R_\oplus$, $P \approx 8.06$~d, $\rho_{\rm p} < 0.48$ g cm$^{-3}$) and an inner small-size companion ($R_{\rm p} = 2.89 \pm 0.06 R_\oplus$, $P \approx 3.65$~d). Despite their disparate radii, these planets have a large mass ratio ($M_{\rm b}/M_{\rm c} \approx 0.8$), and the gas giant has a low mass ($M_{\rm c} \approx 40M_\oplus$), making this system quite unusual compared to classical HJs with inner low-mass companions.
    \item Jointly modelling HARPS-N RVs and \textit{TESS} photometry using a multi-GP framework was fundamental to disentangling the stellar activity from the planetary signals. The inclusion of \textit{TESS} sector 84 in the GP formalism allowed us to halve the jitter.
    \item We detected possible TTVs for planet TOI-1533~b, with a reduced $\chi^2 \approx 7$, which suggests ongoing dynamical interactions. However, significant scatter in the O-C diagram residuals -- likely due to in-transit stellar activity in the form of stellar spots -- needs to be modelled out using future higher-precision photometry, such as that provided by CHEOPS.
    \item We found a total of eleven systems hosting hot Neptunes and Saturns ($15~M_\oplus < M_{\rm p} < 120~M_\oplus$, $P < 20$~d) with inner companions of any kind. None of these is hosted by metal-poor stars, and those orbiting F-G dwarf stars tend to be more massive than those orbiting K dwarfs.
    \item Systems with small planets orbiting interior to a close-in giant often cluster near mean-motion resonances, with gravitational interactions leading to detectable TTVs. The resonant configuration may be disrupted by mass loss or different tidal interactions with the host star, causing the inner companion to detach and increasing the planet-pair period ratio ($\mathcal{P}$). We inspected the eleven systems mentioned above and found a possible correlation between $\mathcal{P}$ and the inner planet period ($P_1$), which supports recent findings by \cite{2025AJ....169..191G}.
\end{itemize}

\begin{acknowledgements}

We acknowledge the use of public TESS data from pipelines at the TESS Science Office and at the TESS Science Processing Operations Center. Funding for the TESS mission is provided by NASA's Science Mission Directorate. This research has made use of the Exoplanet Follow-up Observation Program (ExoFOP; DOI: 10.26134/ExoFOP5) website and the NASA Exoplanet Archive, both of which are operated by the California Institute of Technology under contract with the National Aeronautics and Space Administration under the Exoplanet Exploration Program. S.W.Y. gratefully acknowledges support from the Heising-Simons Foundation. This paper contains data taken with the NEID instrument, which was funded by the NASA-NSF Exoplanet Observational Research (NN-EXPLORE) partnership and built by Pennsylvania State University. Data presented herein were obtained at the WIYN Observatory from telescope time allocated to NN-EXPLORE (2025A-381524, 2025B-123953, PI: Yee) through the scientific partnership of the National Aeronautics and Space Administration, the National Science Foundation, and NOIRLab. The authors are honored to be permitted to conduct astronomical research on Iolkam Du’ag (Kitt Peak), a mountain with particular significance to the Tohono O’odham. This work makes use of observations from the LCOGT network. Part of the LCOGT telescope time was granted by NOIRLab through the Mid-Scale Innovations Program (MSIP). MSIP is funded by NSF. This paper is based on observations made with the Las Cumbres Observatory’s education network telescopes that were upgraded through generous support from the Gordon and Betty Moore Foundation. KAC acknowledges support from the TESS mission via subaward s3449 from MIT and NASA grants 80NSSC24K1889 and 80NSSC26K0081. This paper is based on observations made with the MuSCAT3 instrument, developed by the Astrobiology Center and under financial supports by JSPS KAKENHI (JP18H05439) and JST PRESTO (JPMJPR1775), at Faulkes Telescope North on Maui, HI, operated by the Las Cumbres Observatory. This article is based on observations made with the MuSCAT2 instrument, developed by ABC, at Telescopio Carlos Sánchez operated on the island of Tenerife by the IAC in the Spanish Observatorio del Teide. This work is partly supported by JSPS KAKENHI Grant Numbers JP24H00017, JP24K17083, JP24K00689, JP21K13955, JP24K17082, JP24H00248, JP25K24620, JP26H01402, JP26K00755, and JSPS Grant-in-Aid for JSPS Fellows Grant Number JP24KJ0241. We acknowledge financial support from the Agencia Estatal de Investigación of the Ministerio de Ciencia e Innovación MCIN/AEI/10.13039/501100011033 and the ERDF “A way of making Europe” through projects PID2021-125627OB-C32 and PID2024-158486OB-C32. JK acknowledges support from the Swedish Research Council (Starting Grant 2017-04945 and Project Grant 2022-04043) and from the Spanish Research Agency of the Ministry of Science, Innovation and Universities (AEI-MICIU) under grant 'Contribution of the IAC to the PLATO Space Mission' with reference PID2023-149439NB-C41. F. M. acknowledges the financial support from the Agencia Estatal de Investigaci\'{o}n del Ministerio de Ciencia, Innovaci\'{o}n y Universidades (MCIU/AEI) through grant PID2023-152906NA-I00. LBo, VNa, GPi, PLe acknowledge the funding support from Italian Space Agency (ASI) regulated by ‘Accordo ASI-INAF n. 2013-016-R.0 del 9 luglio 2013 e integrazione del 9 luglio 2015 CHEOPS Fasi A/B/C’. We acknowledge the Italian center for Astronomical Archives (IA2, \url{https://www.ia2.inaf.it}), part of the Italian National Institute for Astrophysics (INAF), for providing technical assistance, services and supporting activities of the GAPS collaboration. JJL was supported by NASA's Exoplanets Research Program grant 24-XRP24\_2-0020.
\end{acknowledgements}

%
\bibliographystyle{aa} 
\bibliography{references} 

\begin{appendix}
\section{Detailed stellar activity modelling}
\label{app:gp}
Our corrected \textit{TESS} light curve for sector 85 displayed significant instrumental variability (mainly stray light). This behaviour often occurred when the photometry was associated with high local background values and forced us to mask these points, making it impossible to get a complete, well-detrended light curve. This impacted our ability to preserve the stellar variability and hence model its short-term properties precisely. 

For this reason, and also because the zero point with respect to sector 84 was challenging to retrieve, we decided not to use \textit{TESS} sector 85 as one of the observables in the multidimensional GP. Nevertheless, we included it in our joint fit and modelled it using a unidimensional GP that shared the stellar rotation period (a rather long-term property) with the multidimensional GP, as done for the other \textit{TESS} sectors.

When we initially attempted to include sector 85 as an observable in the multidimensional GP framework, the Bayesian analysis failed to converge. The reason for this outcome was the difficulty of finding a set of hyperparameters that could consistently explain sector 85 (and its instrumental variability) plus all other \textit{TESS} sectors and activity indicators. 

\section{Dynamical analysis}
\label{app:dyn}
The retrieved transit times ($T_0$) of TOI-1533~b and TOI-1533~c were modelled by numerically integrating the system with the N-body integrator TRADES. We adopted the same fitting framework, the same analysis procedure (PyDE exploration, $\texttt{emcee}$ sampling, and convergence criteria) described in \citet{2025A&A...702A.211L}. Specifically, we used an integration time ($T_\textrm{int}$) of 2506 days that covers the entire observation time span, and a time start of the integration $T_\textrm{ref, dyn} = 2458760.0 ~\textrm{BJD}_\textrm{TDB}$.

This analysis successfully retrieved planetary parameters that are consistent with those reported in Table \ref{table:model-lcrv}. However, there is still a significant scatter in the residuals of both O-C diagrams. The most likely source of this scatter is in-transit stellar activity affecting the \textit{TESS} light-curves. TTVs of about 10 minutes could be explained by the presence of stellar spots \citep[e.g.][]{2026AJ....171..344M}, but these could only be modelled out of the transits with exquisite, high-precision photometry, such as CHEOPS.

\section{Independent stellar parameters determination}
 As an independent determination of the basic stellar parameters, we performed an analysis of the broadband spectral energy distribution (SED) of the star together with the {\it Gaia\/} DR3 parallax \citep[with no systematic offset applied; see e.g.][]{StassunTorres:2021}, to determine an empirical measurement of the stellar radius, following the procedures described in \citet{Stassun:2016,2017AJ....153..136S,Stassun:2018}. We pulled the $JHK_S$ magnitudes from {\it 2MASS}, the $G_{\rm BP} G_{\rm RP}$ magnitudes from {\it Gaia}, and the W1--W4 magnitudes from {\it WISE}. We also utilised the absolute flux-calibrated spectrophotometry from {\it Gaia}. Together, the available photometry spans the full stellar SED over the wavelength range 0.4--20~$\mu$m (see Figure~\ref{fig:sed}).  
 
We performed a fit using PHOENIX stellar atmosphere models \citep{Husser:2013}, with the effective temperature ($T_{\rm eff}$), surface gravity ($\log g$), and metallicity ([Fe/H]) adopted from the spectroscopic analysis. The extinction, $A_V$, was limited to the maximum line-of-sight value from the Galactic dust maps of \citet{Schlegel:1998}. The resulting fit (Figure~\ref{fig:sed}) has a reduced $\chi^2$ of 1.5, with a best-fit $A_V = 0.01 \pm 0.01$. Integrating the (unreddened) model SED gives the bolometric flux at Earth, $F_{\rm bol} = 1.396 \pm 0.016 \times 10^{-9}$ erg~s$^{-1}$~cm$^{-2}$. Taking the $F_{\rm bol}$ together with the {\it Gaia\/} parallax directly gives the bolometric luminosity, $L_{\rm bol} = 0.4349 \pm 0.0051$~L$_\odot$. The stellar radius follows from the Stefan-Boltzmann relation, giving $R_\star = 0.837 \pm 0.024$~R$_\odot$. In addition, we can estimate the stellar mass from the empirical relations of \citet{Mann:2019}, giving $M_\star = 0.82 \pm 0.04$~M$_\odot$.

\begin{figure}
   \centering
   \includegraphics[trim={3.5cm 2.5cm 2cm 2.5cm},width=\hsize]%
   {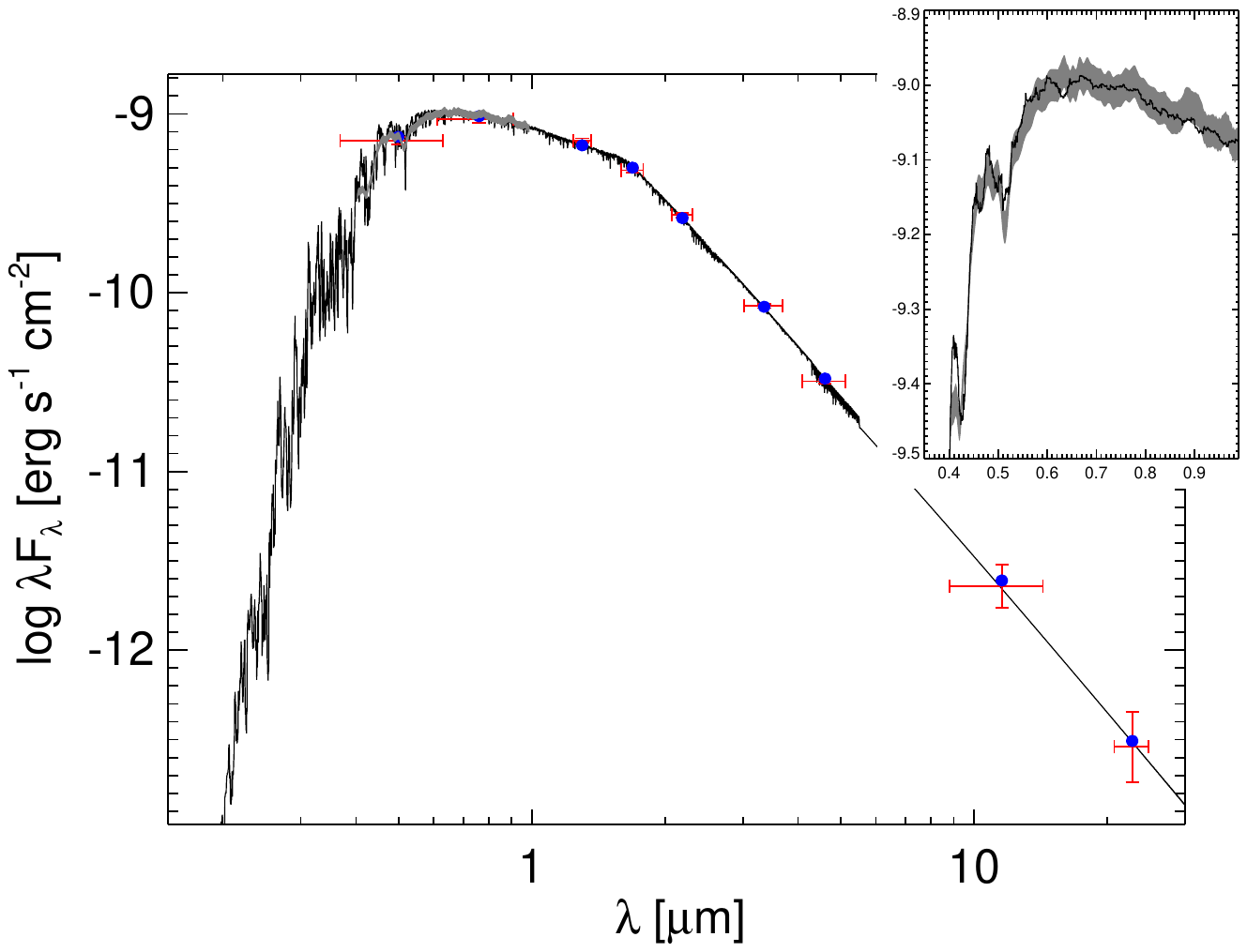}
   \caption{Spectral energy distribution of TOI-1533. Red symbols represent the observed photometric measurements, where the horizontal bars represent the effective width of the passband. Blue symbols are the model fluxes from the best-fit PHOENIX atmosphere model (black). The {\it Gaia\/} spectrophotometry is overlaid as the grey swathe, also in the inset plot. \label{fig:sed}}
   \label{fig:sed}
\end{figure}

\section{Further atmospheric follow-up prospects}
\label{app:further}
The expected day-to-night temperature contrasts \citep{2016ApJ...821...16K} induce longitudinal chemical gradients \citep{2023MNRAS.519.3129Z}, leading to compositional asymmetries between the eastern and western limbs \citep{2024Natur.632.1017E}, given the equilibrium temperature of this planet. However, the grazing nature of the transit makes TOI-1533 c a challenging target to interpret. The transit geometry reduces the effective atmospheric annulus probed in transmission and limits the fraction of the terminator that is sampled. At the same time, this partial sampling presents a unique opportunity, as it may enhance sensitivity to localised atmospheric regions rather than averaging uniformly over the full limb. Such a geometry could enable novel constraints on wind patterns and limb asymmetries, particularly when combined with ground-based high-resolution spectroscopy capable of resolving Doppler-shifted molecular lines. Measurements of line shifts and broadening could, in principle, isolate contributions from specific regions of the terminator. However, the lack of constraints on the planet’s three-dimensional orientation and obliquity complicates the interpretation of the observed spectra through transmission spectroscopy. Although difficult, a detailed atmospheric characterisation of this system would provide a rare opportunity to link disk chemistry, dynamical evolution, and present-day atmospheric composition within a single planetary architecture.

\section{Hot giants and intermediate-mass planets with inner companions}

\begin{table}[t]
\tiny
\caption{List of systems presented in this work.}  
\label{tab:companions}
\centering   
\addtolength{\tabcolsep}{-0.45em}
\begin{tabular}{c | c c c}     
\hline\hline         
 System & Class\tablefootmark{a} & Discovery paper & Parameters paper \rule{0pt}{2.5ex} \rule[-1ex]{0pt}{0pt} \\
\hline         
    WASP-47 &  H & \citet{2015ApJ...812L..18B} & \citet{2023AA...673A..42N}  \rule{0pt}{2.5ex} \rule[-1ex]{0pt}{0pt} \\
    TOI-1408  & H & \citet{2024ApJ...971L..28K} & Same as discovery \rule{0pt}{2.5ex} \rule[-1ex]{0pt}{0pt} \\
    TOI-5143  & H & \citet{2025AJ....169..189R} & / \rule{0pt}{2.5ex} \rule[-1ex]{0pt}{0pt} \\
    Kepler-730 & H & \citet{2019ApJ...870L..17C} & / \rule{0pt}{2.5ex} \rule[-1ex]{0pt}{0pt} \\
    WASP-132  & H & \citet{2022AJ....164...13H} & \citet{2025AA...693A.144G} \rule{0pt}{2.5ex} \rule[-1ex]{0pt}{0pt} \\
    Kepler-18 & IM & \citet{2011ApJS..197....7C} & Same as discovery \rule{0pt}{2.5ex} \rule[-1ex]{0pt}{0pt} \\
    Kepler-411 & IM & \citet{2014ApJ...783....4W} & \citet{2019AA...624A..15S} \rule{0pt}{2.5ex} \rule[-1ex]{0pt}{0pt} \\
    K2-19 & IM & \citet{2015MNRAS.454.4267B} & \citet{2025AA...703A.167A} \rule{0pt}{2.5ex} \rule[-1ex]{0pt}{0pt} \\
    TOI-1533 & IM & This work & This work \rule{0pt}{2.5ex} \rule[-1ex]{0pt}{0pt} \\
    TOI-1130  & H & \citet{Huang_2020} & \citet{2024AA...689A..52B} \rule{0pt}{2.5ex} \rule[-1ex]{0pt}{0pt} \\
    TOI-1272 & IM & \citet{2022AJ....164...97M} & Same as discovery \rule{0pt}{2.5ex} \rule[-1ex]{0pt}{0pt} \\
    K2-32 & IM & \citet{2016ApJ...823..115D} & \citet{2020AA...640A..48L}\rule{0pt}{2.5ex} \rule[-1ex]{0pt}{0pt} \\
    TOI-2000  & H, IM & \citet{2023MNRAS.524.1113S} & Same as discovery \rule{0pt}{2.5ex} \rule[-1ex]{0pt}{0pt} \\
    TOI-5398  & H, IM & \citet{2022MNRAS.516.4432M} & \citet{2024AA...682A.129M} \rule{0pt}{2.5ex} \rule[-1ex]{0pt}{0pt} \\
    CoRoT-24 & IM & \citet{2014AA...567A.112A} & Same as discovery \rule{0pt}{2.5ex} \rule[-1ex]{0pt}{0pt} \\
    Kepler-25 & IM & \citet{2012MNRAS.421.2342S} & \citet{2019AJ....157..145M} \rule{0pt}{2.5ex} \rule[-1ex]{0pt}{0pt} \\
    TOI-4010 & IM & \citet{2023AJ....166....7K} & Same as discovery \rule{0pt}{2.5ex} \rule[-1ex]{0pt}{0pt} \\
    Kepler-9 & IM & \citet{2010Sci...330...51H} & \citet{2019MNRAS.484.3233B} \rule{0pt}{2.5ex} \rule[-1ex]{0pt}{0pt} \\
\hline      
\hline                                   
\end{tabular}
\tablefoot{\tablefoottext{a}{System class: H stands for `Hot giant' and IM stands for `intermediate-mass' planet.}}
\end{table}

\section{Extra figures and tables}

\begin{figure}[h]
   \centering
   \includegraphics[width=\hsize]%
   {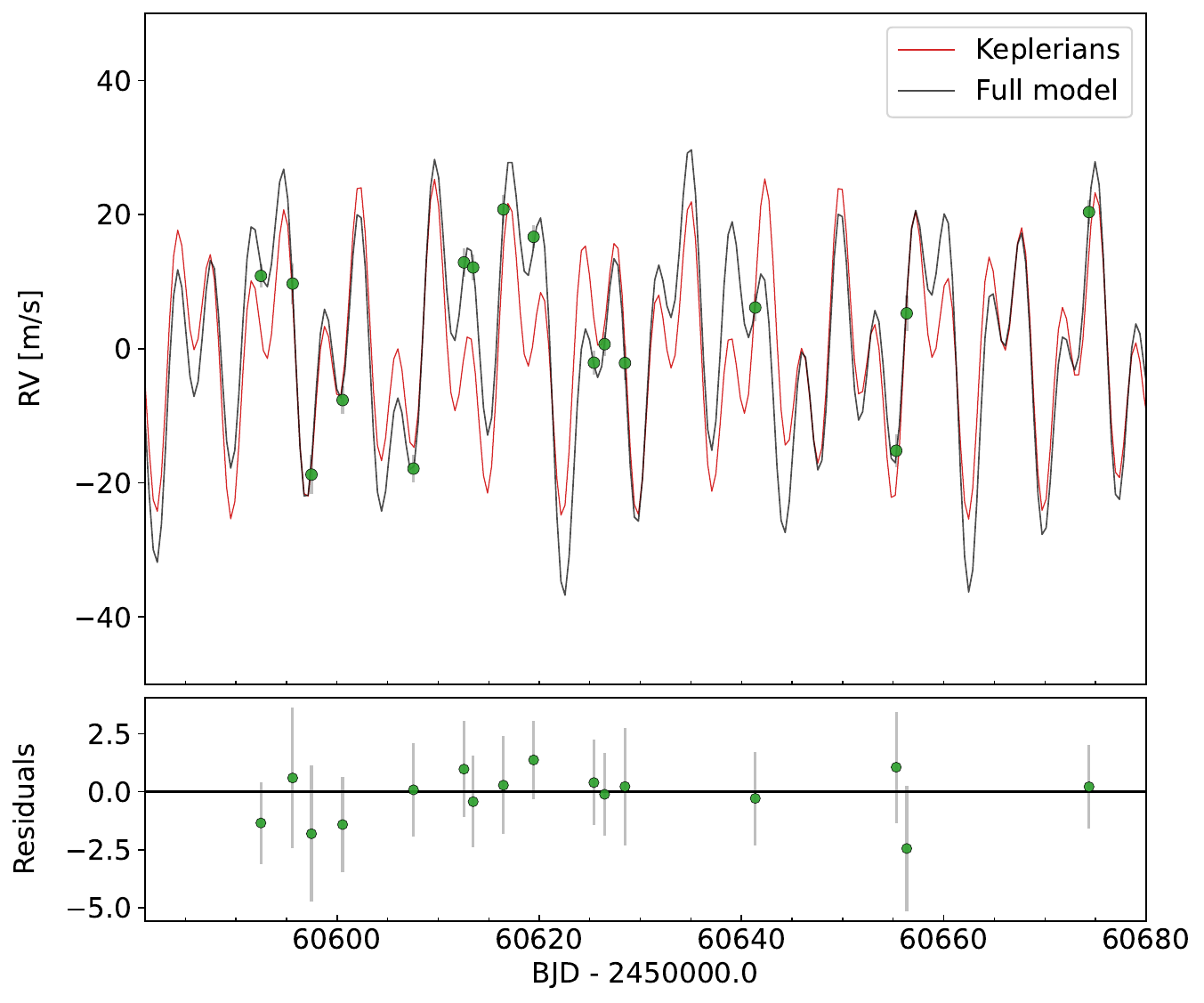}
   \caption{Modelling of HARPS-N RV time series. \textit{Top:} RV time series with superimposed full Keplerian $+$ stellar activity model (black line), and full Keplerian only (red line). \textit{Bottom:} Residuals of the fit.}
   \label{fig:rv_full}
\end{figure}

\begin{figure}[h]
   \centering
   \includegraphics[width=\hsize]%
   {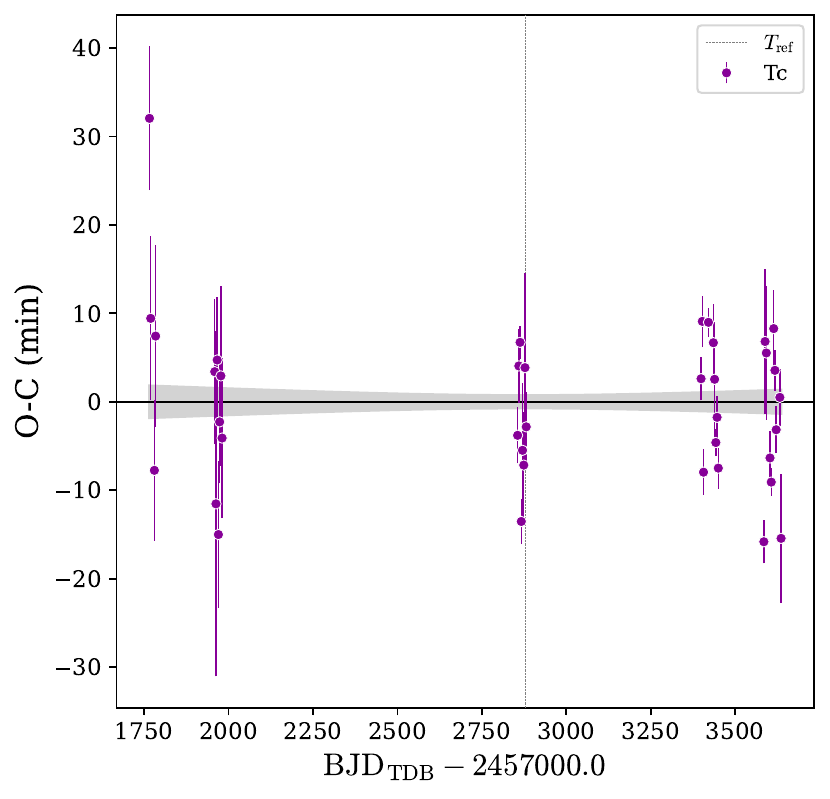}
   \caption{Observed minus calculated transit times for the linear ephemeris of TOI-1533~b. The grey area shows the formal uncertainty of the linear ephemeris.}
   \label{fig:ttv_b}
\end{figure}

\begin{figure}[h]
   \centering
   \includegraphics[width=\hsize]%
   {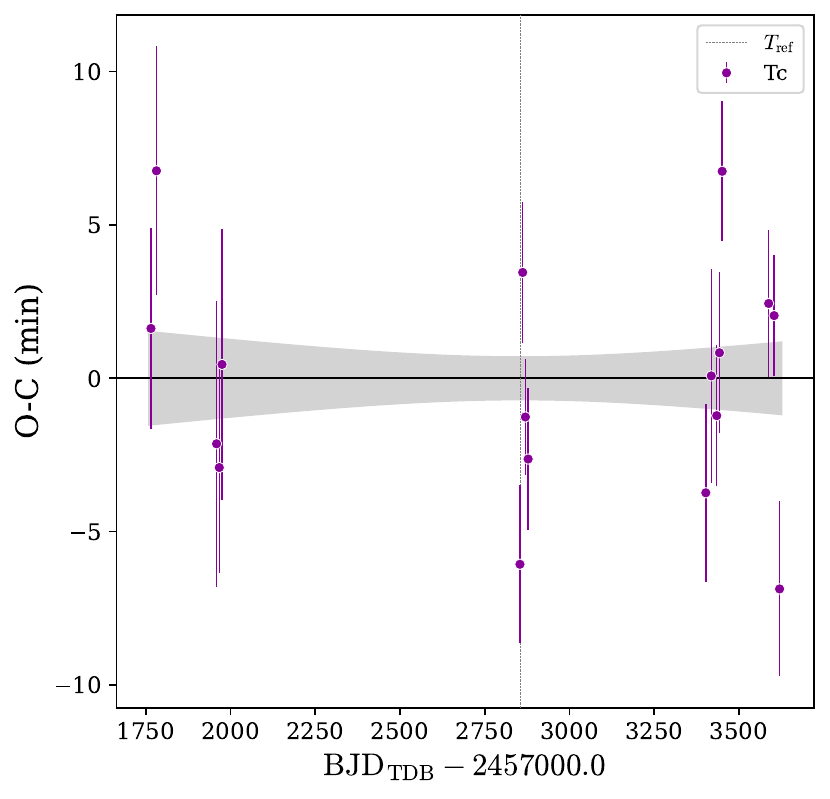}
   \caption{Same as in figure \ref{fig:ttv_b} but for planet TOI-1533~c.}
   \label{fig:ttv_c}
\end{figure}

\begin{figure}[h]
   \centering
   \includegraphics[width=0.9\hsize]%
   {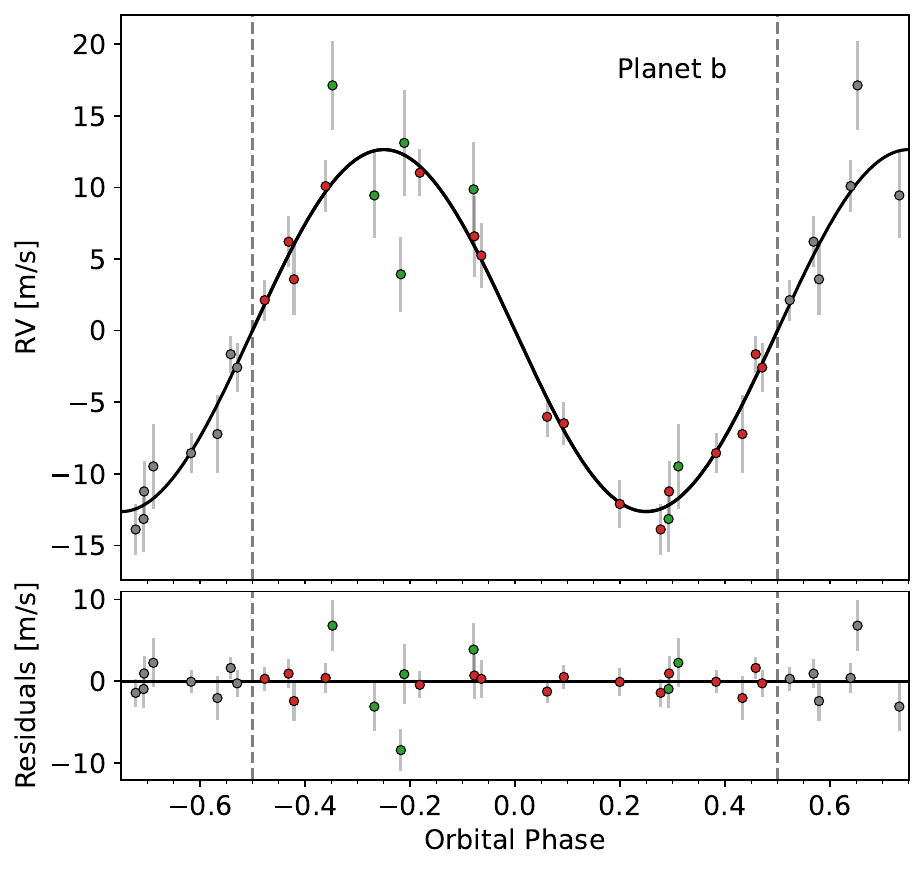}
   \caption{Phase-folded RV curve of TOI-1533 b with NEID RVs included. The different colours represent different datasets: red for HARPS-N and green for NEID. The residuals of the fit are shown in the \textit{bottom} panel.}
   \label{fig:1533rvb_neid}
\end{figure}

\begin{figure}[h]
   \centering
   \includegraphics[width=0.9\hsize]%
   {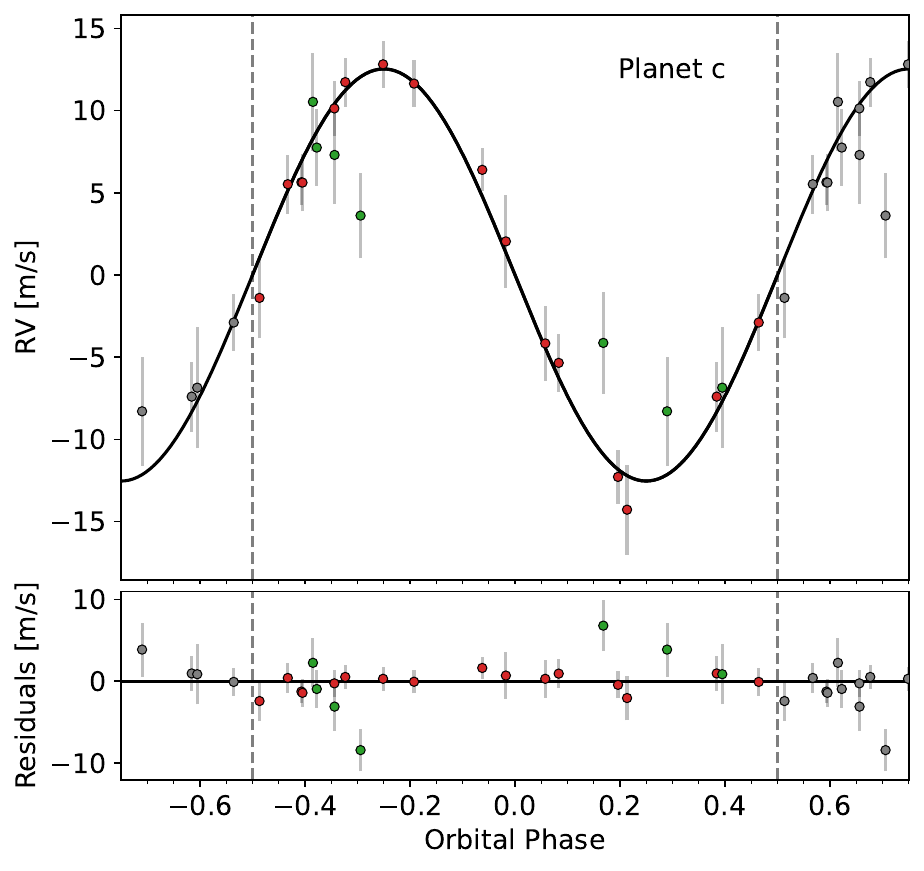}
   \caption{Same as in Figure \ref{fig:1533rvb_neid} but for planet TOI-1533~c.}
   \label{fig:1533rvc_neid}
\end{figure}

\begin{figure}[h]
   \centering
   \includegraphics[width=\hsize]%
   {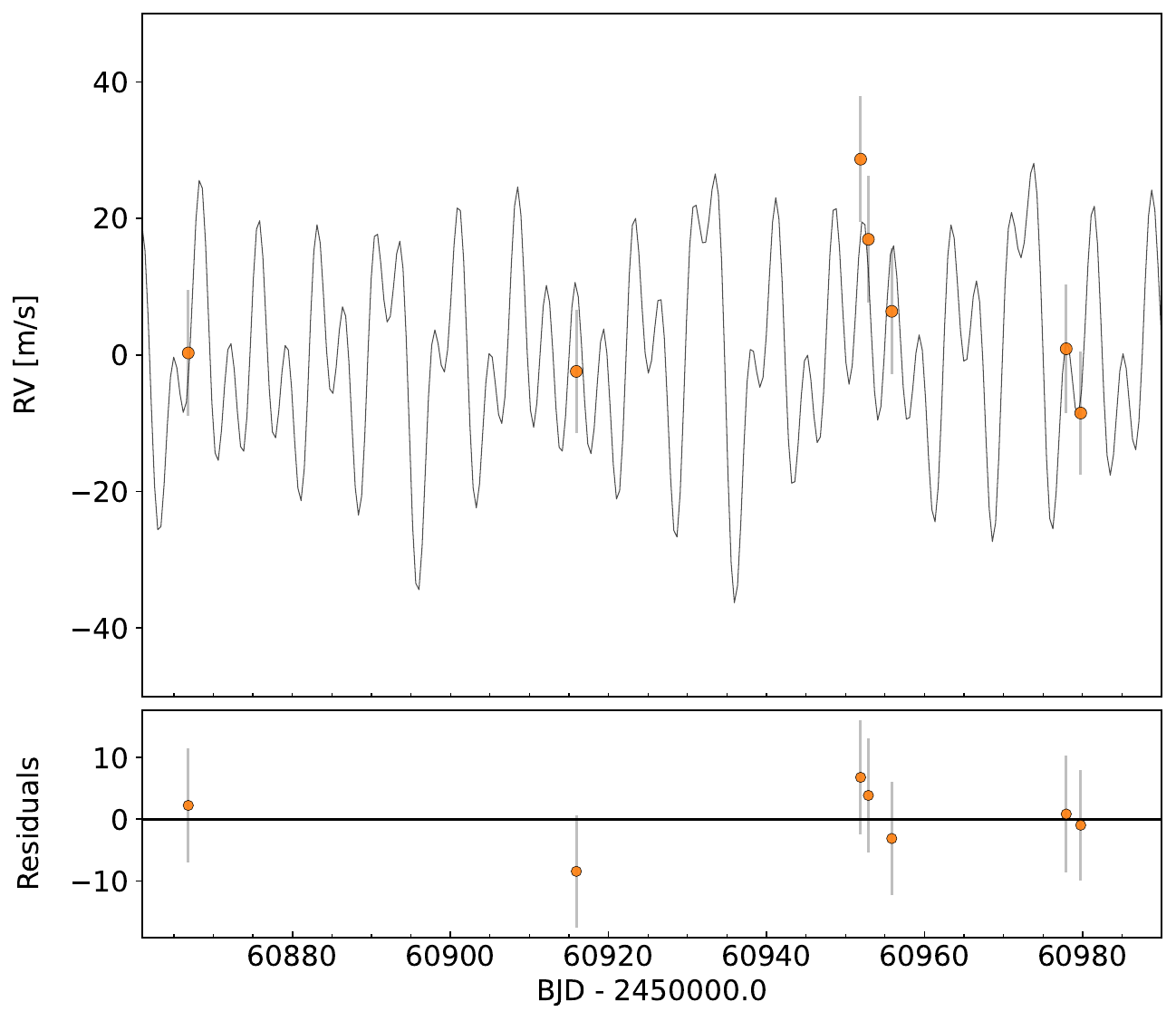}
   \caption{Same as in Figure \ref{fig:rv_full} but for NEID data.}
   \label{fig:rv_full_neid}
\end{figure}

\begin{figure}
   \centering
   \includegraphics[width=0.75\hsize]{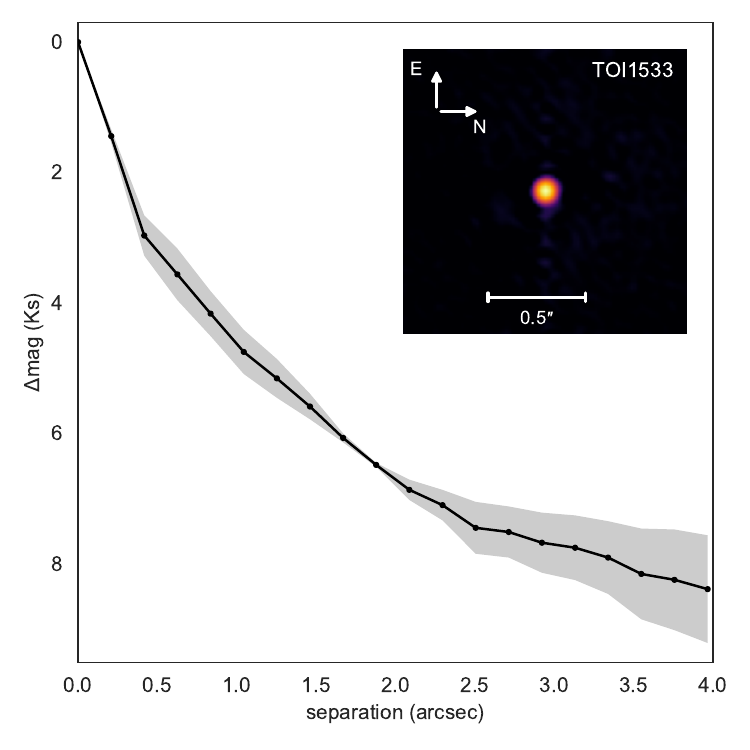}
   \caption{Adaptive optics images of TOI-1533 taken with the ShARCS camera. We also present a contrast curve generated by calculating the median values (solid lines) and root-mean-square errors (blue, shaded regions) in annuli centred on each target, where the bin width of each annulus is equal to the full width at half max of the point spread function.}
   \label{fig:sharcs}
\end{figure}

\begin{figure}
   \centering
   \includegraphics[width=1\hsize]{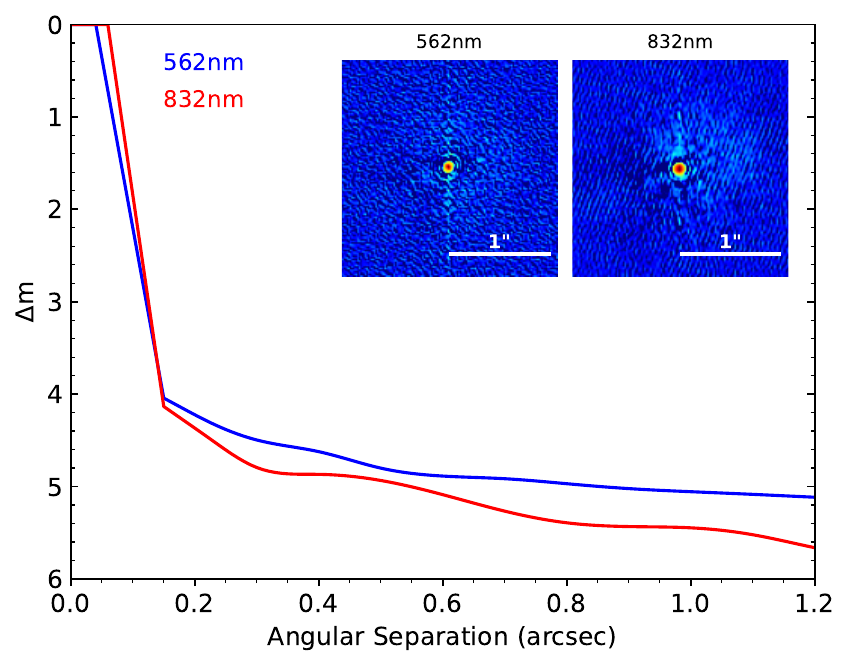}
   \caption{Contrast curves for TOI-1533 at 562~nm (blue) and 832~nm (red) obtained from NESSI speckle imaging. Reconstructed images are shown as insets to the plot with wavelengths labelled. Here, North is towards the top and East to the left.}
   \label{fig:nessi}
\end{figure}

\begin{figure}
   \centering
   \includegraphics[width=\hsize]%
   {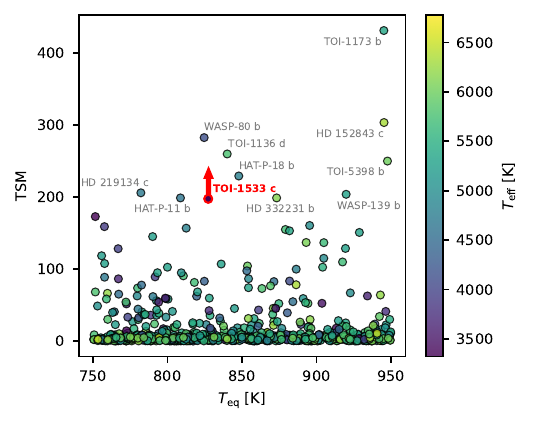}
   \caption{Transmission spectroscopy metric for planets with equilibrium temperatures between 750 and 850 K. TOI-1533 c is highlighted in red, and its TSM was computed assuming the minimum reported planetary radius. Points are colour-coded by the effective temperature of the host star. Planetary and stellar parameters are from the NASA Exoplanet Archive \citep{2025PSJ.....6..186C}.}
   \label{fig:TSM_TOI_1533c}
\end{figure}

\begin{figure}
   \centering
   \includegraphics[width=0.8\hsize]%
   {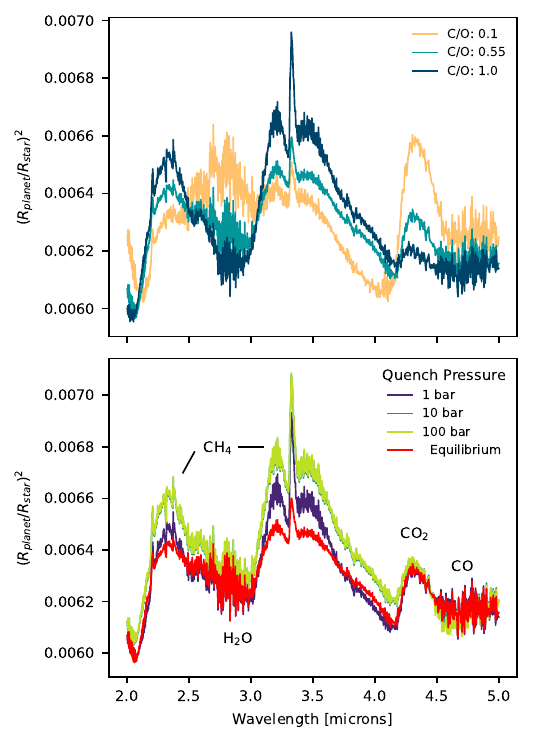}
   \caption{Simulated transmission spectrum of TOI-1533 c generated with petitRADTRANS \citep{2019A&A...627A..67M} using the planetary parameters reported in this work and a Guillot temperature–pressure profile \citep[][$\kappa_{\mathrm{IR}} = 0.05$, $\gamma = 0.1$, and $T_{\mathrm{int}} = 100\,\mathrm{K}$]{2010A&A...520A..27G}. Top panel: Spectra for different atmospheric carbon-to-oxygen (C/O) ratios. Bottom panel: Comparison between chemical equilibrium and disequilibrium models with varying quench pressures, assuming C/O = 0.55.}
   \label{fig:simulated_spectra_TSM_TOI_1533c}
\end{figure}

\begin{table}
\caption{Priors and outcomes of full modelling.}          
\label{table:model-full}
\addtolength{\tabcolsep}{-0.7em}
\centering          
\begin{tabular}{l c c c c} 
\hline\hline     
Parameter & Unit & Prior & Value \rule{0pt}{2.3ex} \rule[-1ex]{0pt}{0pt}\\ 
\hline 
   \textit{TESS} s17 jitter ($\sigma^{\rm s17}_{jitter}$) & ppt & ... & 0.29$\pm$0.01 \rule{0pt}{2.3ex} \rule[-1ex]{0pt}{0pt}\\
   \textit{TESS} s24 jitter ($\sigma^{\rm s24}_{jitter}$) & ppt & ... & 0.34$\pm$0.02 \rule{0pt}{2.3ex} \rule[-1ex]{0pt}{0pt}\\
   \textit{TESS} s57 jitter ($\sigma^{\rm s57}_{jitter}$) & ppt & ... & 0.19$\pm$0.01 \rule{0pt}{2.3ex} \rule[-1ex]{0pt}{0pt}\\
   \textit{TESS} s77 jitter ($\sigma^{\rm s77}_{jitter}$) & ppt & ... & 0.12$\pm$0.02 \rule{0pt}{2.3ex} \rule[-1ex]{0pt}{0pt}\\
   \textit{TESS} s78 jitter ($\sigma^{\rm s78}_{jitter}$) & ppt & ... & 0.26$^{+0.03}_{-0.02}$ \rule{0pt}{2.3ex} \rule[-1ex]{0pt}{0pt}\\
   \textit{TESS} s84 jitter ($\sigma^{\rm s84}_{jitter}$) & ppt & ... & 0.23$\pm$0.02 \rule{0pt}{2.3ex} \rule[-1ex]{0pt}{0pt}\\
   \textit{TESS} s85 jitter ($\sigma^{\rm s85}_{jitter}$) & ppt & ... & 0.15$\pm$0.02 \rule{0pt}{2.3ex} \rule[-1ex]{0pt}{0pt}\\
   \textit{MuSCAT2} jitter ($\sigma^{\rm MuSCAT2}_{jitter}$) & ppt & ... & 0.32$\pm$0.20 \rule{0pt}{2.3ex} \rule[-1ex]{0pt}{0pt}\\
   Teide 23/12/03 jitter ($\sigma^{\rm Teide}_{jitter}$) & ppt & ... & 0.53$\pm$0.10 \rule{0pt}{2.3ex} \rule[-1ex]{0pt}{0pt}\\
   Teide 24/08/04 jitter ($\sigma^{\rm Teide}_{jitter}$) & ppt & ... & 0.53$\pm$0.13 \rule{0pt}{2.3ex} \rule[-1ex]{0pt}{0pt}\\
   Teide 24/08/11 jitter ($\sigma^{\rm Teide}_{jitter}$) & ppt & ... & 1.16$\pm$0.10 \rule{0pt}{2.3ex} \rule[-1ex]{0pt}{0pt}\\
   Hal 23/10/07 jitter ($\sigma^{\rm Hal}_{jitter}$) & ppt & ... & 1.40$\pm$0.18 \rule{0pt}{2.3ex} \rule[-1ex]{0pt}{0pt}\\
   Hal 24/10/12 jitter ($\sigma^{\rm Hal}_{jitter}$) & ppt & ... & 0.81$\pm$0.07 \rule{0pt}{2.3ex} \rule[-1ex]{0pt}{0pt}\\
   Hal 25/11/03 jitter ($\sigma^{\rm Hal}_{jitter}$) & ppt & ... & 0.65$\pm$0.04 \rule{0pt}{2.3ex} \rule[-1ex]{0pt}{0pt}\\
Uncorr. RHK jitter ($\sigma^{\rm RHK}_{jitter}$) &  & ... & 0.004$^{+0.004}_{-0.003}$   \rule{0pt}{2.3ex} \rule[-1ex]{0pt}{0pt}\\
Uncorr. FWHM jitter & km s$^{-1}$ & ... & 0.002$\pm$0.002   \rule{0pt}{2.3ex} \rule[-1ex]{0pt}{0pt}\\
RHK offset ($\gamma^{\rm RHK}$) &  & ... & $-$4.49$^{+0.01}_{-0.02}$   \rule{0pt}{2.3ex} \rule[-1ex]{0pt}{0pt}\\
FWHM offset ($\gamma^{\rm FWHM}$) & km s$^{-1}$ & ... & 6.74$^{+0.01}_{-0.02}$   \rule{0pt}{2.3ex} \rule[-1ex]{0pt}{0pt}\\
\textit{TESS} quad. LD coeff. ($u_1$) &  & $\mathcal{U}$(0, 1) & 0.38$^{+0.33}_{-0.26}$ \rule{0pt}{2.3ex} \rule[-1ex]{0pt}{0pt}\\
\textit{TESS} quad. LD coeff. ($u_2$) &  & $\mathcal{U}$(0, 1) & 0.24$^{+0.30}_{-0.38}$ \rule{0pt}{2.3ex} \rule[-1ex]{0pt}{0pt}\\
\textit{MuSCAT2 $i'$} LD ($u_1$) &  & $\mathcal{N}$(0.50, 0.10) & 0.54$\pm$0.08 \rule{0pt}{2.3ex} \rule[-1ex]{0pt}{0pt}\\
\textit{MuSCAT2 $i'$} LD ($u_2$) &  & $\mathcal{N}$(0.11, 0.11) & 0.16$\pm$0.09 \rule{0pt}{2.3ex} \rule[-1ex]{0pt}{0pt}\\
\textit{LCO Sinistro z$_s$} LD ($u_1$) &  & $\mathcal{N}$(0.42, 0.10) & 0.43$\pm$0.09 \rule{0pt}{2.3ex} \rule[-1ex]{0pt}{0pt}\\
\textit{LCO Sinistro z$_s$} LD ($u_2$) &  & $\mathcal{N}$(0.12, 0.11) & 0.13$\pm$0.10 \rule{0pt}{2.3ex} \rule[-1ex]{0pt}{0pt}\\
\textit{MuSCAT3 $g'$} LD ($u_1$) &  & $\mathcal{N}$(0.84, 0.10) & 0.82$^{+0.07}_{-0.08}$ \rule{0pt}{2.3ex} \rule[-1ex]{0pt}{0pt}\\
\textit{MuSCAT3 $g'$} LD ($u_2$) &  & $\mathcal{N}$($-$0.02, 0.11) & $-$0.04$^{+0.09}_{-0.08}$ \rule{0pt}{2.3ex} \rule[-1ex]{0pt}{0pt}\\
   \textit{GP parameters} & & &  \rule{0pt}{2.3ex} \rule[-1ex]{0pt}{0pt}\\
Rot$_{\rm Q0}$ &  & ... & 0.78$^{+0.55}_{-0.32}$ \rule{0pt}{2.3ex} \rule[-1ex]{0pt}{0pt}\\
Rot$_{\rm deltaQ}$ &  & ... & 0.0001$^{+0.08}_{-0.0001}$ \rule{0pt}{2.3ex} \rule[-1ex]{0pt}{0pt}\\
Rot$_{\rm fmix}$ &  & ... & 0.128$^{+0.25}_{-0.099}$ \rule{0pt}{2.3ex} \rule[-1ex]{0pt}{0pt}\\
Rot$_\sigma$ s17 &  & ... & 0.0064$^{+0.002}_{-0.001}$ \rule{0pt}{2.3ex} \rule[-1ex]{0pt}{0pt}\\
Rot$_\sigma$ s24 &  & ... & 0.0058$^{+0.0012}_{-0.0009}$ \rule{0pt}{2.3ex} \rule[-1ex]{0pt}{0pt}\\
Rot$_\sigma$ s57 &  & ... & 0.0062$^{+0.0011}_{-0.0009}$ \rule{0pt}{2.3ex} \rule[-1ex]{0pt}{0pt}\\
Rot$_\sigma$ s77 &  & ... & 0.0033$^{+0.0007}_{-0.0005}$ \rule{0pt}{2.3ex} \rule[-1ex]{0pt}{0pt}\\
Rot$_\sigma$ s78 &  & ... & 0.014$^{+0.003}_{-0.002}$ \rule{0pt}{2.3ex} \rule[-1ex]{0pt}{0pt}\\
Rot$_\sigma$ s85 &  & ... & 0.009$\pm$0.002 \rule{0pt}{2.3ex} \rule[-1ex]{0pt}{0pt}\\
$V_c$ (RV) & m s$^{-1}$ & $\mathcal{U}$(0, 40) & 5.8$^{+2.1}_{-1.5}$   \rule{0pt}{2.3ex} \rule[-1ex]{0pt}{0pt}\\
$V_r$ (RV) & m s$^{-1}$ & $\mathcal{U}$($-$40, 40) & 18.3$^{+5.8}_{-4.3}$   \rule{0pt}{2.3ex} \rule[-1ex]{0pt}{0pt}\\
$L2_c$ (RHK) &  & $\mathcal{U}$($-$0.5, 0.5) & 0.032$^{+0.010}_{-0.007}$   \rule{0pt}{2.3ex} \rule[-1ex]{0pt}{0pt}\\
$L3_c$ (Flux) & ppt & $\mathcal{U}$($-$1000, 1000) & $-$6$^{+1}_{-2}$   \rule{0pt}{2.3ex} \rule[-1ex]{0pt}{0pt}\\
$L_c$ (FWHM) & km s$^{-1}$ & $\mathcal{U}$($-$0.2, 0.2) & 0.033$^{+0.010}_{-0.006}$   \rule{0pt}{2.3ex} \rule[-1ex]{0pt}{0pt}\\
\hline
\end{tabular}
\end{table}

\begin{figure*}[h]
   \centering
   \includegraphics[width=\hsize]%
   {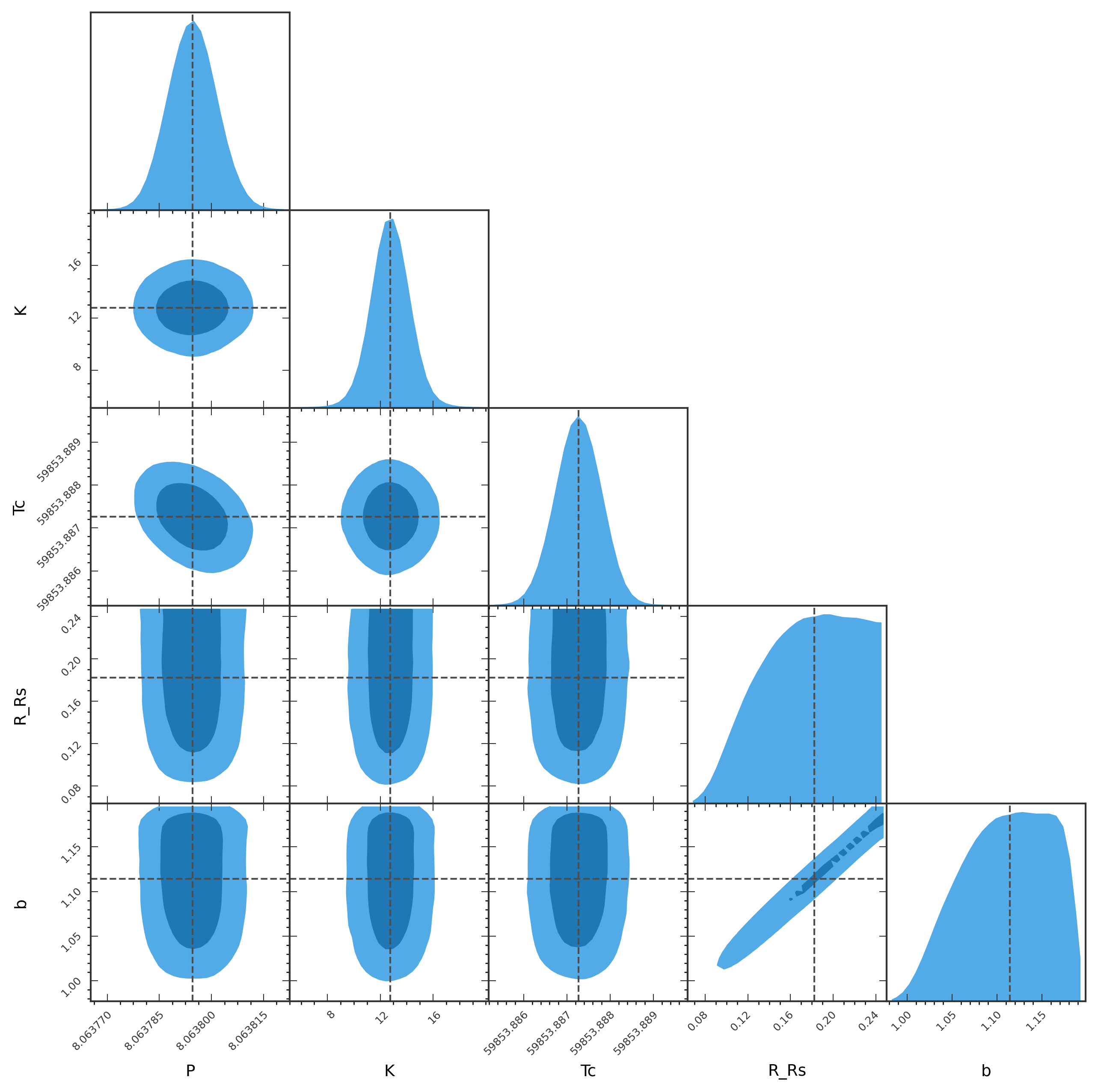}
   \caption{Posterior distributions of TOI-1533~c transit model fit.}
   \label{fig:post_distr}
\end{figure*}

\end{appendix}
\end{document}